\documentclass[fleqn,usenatbib]{mnras}

\usepackage{newtxtext,newtxmath}
\usepackage[T1]{fontenc}
\usepackage{makecell}
\usepackage{soul}
\usepackage{gensymb}
\usepackage{comment}
\DeclareRobustCommand{\VAN}[3]{#2}
\let\VANthebibliography\thebibliography
\def\thebibliography{\DeclareRobustCommand{\VAN}[3]{##3}\VANthebibliography}

\usepackage{graphicx}	% Including figure files
\usepackage{amsmath}	% Advanced maths commands
\newcommand{\cosmorate}{{\sc cosmo$\mathcal{R}$ate}}
\newcommand{\sevn}{{\sc{sevn}} }
\newcommand{\asloth}{{\sc{a-sloth}} }

\newcommand{\orcidicon}[1]{\href{https://orcid.org/#1}{\includegraphics[width=11pt]{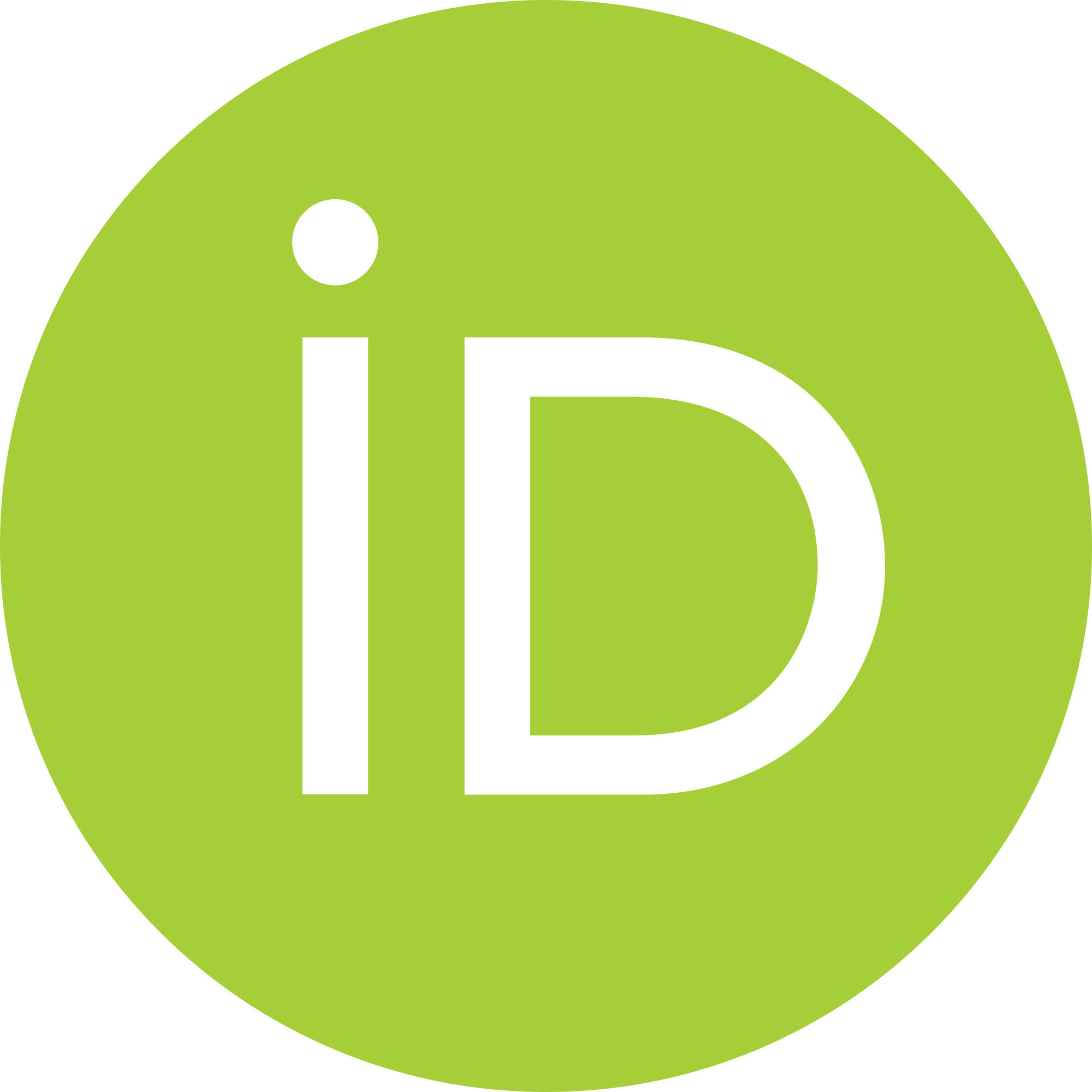}}}

\newcommand{\orcid}[1]{\href{https://orcid.org/#1}{\protect\orcidicon{#1}}}

\defcitealias{giacobbo2020}{GM20}
\defcitealias{kroupa2001}{K01}
\defcitealias{larson1998}{L98}
\defcitealias{sana2012}{S12}
\defcitealias{stacy2013}{SB13}
\defcitealias{hartwig2022}{H22}
\defcitealias{jaacks2019}{J19}
\defcitealias{liu_bromm_20}{LB20}
\defcitealias{skinner2020}{SW20}

\title[BBH mergers from Pop. III stars]{Binary black hole mergers from Population III stars: uncertainties from star formation and binary star properties}

\author[F. Santoliquido et al.]{
 Filippo Santoliquido$^{1,2}$\thanks{E-mail:\href{mailto:filippo.santoliquido@phd.unipd.it}{filippo.santoliquido@unipd.it} (FS)}\orcid{0000-0003-3752-1400}, Michela Mapelli$^{1,2,3}$\thanks{E-mail:\href{mailto:michela.mapelli@unipd.it}{michela.mapelli@unipd.it}}\orcid{0000-0001-8799-2548}, Giuliano Iorio$^{1,2,3}$\thanks{E-mail:\href{mailto:giuliano.iorio@unipd.it}{giuliano.iorio@unipd.it}}\orcid{0000-0003-0293-503X}, Guglielmo Costa$^{1,2,3,4}$\orcid{0000-0002-6213-6988}, \newauthor{Simon C.~O. Glover$^5$\orcid{0000-0001-6708-1317}, Tilman Hartwig$^{6,7,8}$\orcid{0000-0001-6742-8843}, Ralf S. Klessen$^5$\orcid{0000-0002-0560-3172}, and Lorenzo Merli$^1$\orcid{0009-0000-8587-0651}}
 \\
$^{1}$Physics and Astronomy Department Galileo Galilei, University of Padova, Vicolo dell'Osservatorio 3, I--35122, Padova, Italy\\
$^{2}$INFN--Padova, Via Marzolo 8, I--35131, Padova, Italy\\
$^{3}$INAF--Osservatorio Astronomico di Padova, Vicolo dell'Osservatorio 5, I--35122, Padova, Italy\\
$^4$Univ Lyon, Univ Lyon1, Ens de Lyon, CNRS,Centre de Recherche Astrophysique de Lyon UMR5574,  F-69230, Saint-Genis-Laval, France\\
$^{5}$Universit\"at Heidelberg, Zentrum f\"ur Astronomie, Institut f\"ur Theoretische Astrophysik, Albert-Ueberle-Str. 2, D--69120 Heidelberg, Germany \\
$^{6}$Department of Physics, School of Science, The University of Tokyo, Bunkyo, Tokyo 113-0033, Japan\\
$^7$Institute for Physics of Intelligence, School of Science, The University of Tokyo, Bunkyo, Tokyo 113-0033, Japan\\
$^8$Kavli Institute for the Physics and Mathematics of the Universe (WPI), The University of Tokyo Institutes for Advanced Study, The University of Tokyo, Kashiwa, \\Chiba 277-8583, Japan\\
}

\date{Accepted XXX. Received YYY; in original form ZZZ}

\pubyear{2015}

\begin{document}
\label{firstpage}
\pagerange{\pageref{firstpage}--\pageref{lastpage}}
\maketitle

\begin{abstract}
Population~III (Pop.~III) binary stars likely produced the first stellar-born binary black hole (BBH) mergers in the Universe.  Here, we quantify the main sources of uncertainty for the merger rate density evolution and mass spectrum of  Pop.~III BBHs by considering four different  formation histories  %\textcolor{red}{\textbf{
and 11  models of the initial orbital properties of Pop. III binary stars. %}}
The uncertainty on the orbital properties affects the BBH merger rate density by up to two orders of %\textcolor{red}{\textbf{
magnitude, models with shorter orbital periods leading to higher BBH merger rates. %}}
The uncertainty on the star formation history has a substantial impact on both the shape and the normalisation of the BBH merger rate density: the peak of the merger rate density shifts from $z\sim{8}$ up to $z\sim{16}$ depending on the assumed star formation rate, while the maximum BBH merger rate density for our fiducial binary population model spans from $\sim{2}$ to $\sim{30}$ Gpc$^{-3}$ yr$^{-1}$. The typical BBH masses are not affected by the star formation rate model and only mildly influenced by the binary population parameters. The primary black holes born from Pop.~III stars tend to be rather massive ($30-40$ M$_\odot$) with respect to those born from metal-rich stars ($8-10$ M$_\odot$). %\textcolor{red}{\textbf{
We estimate that the Einstein Telescope will detect $10-10^4$ Pop.~III BBH mergers per year, depending on the star formation history and binary star properties.%}}
\end{abstract}


%%%%%%%%%%%%%%%%%%%%%%%%%%%%%%%%%%%%%%%%%%%%%%%%%%

%%%%%%%%%%%%%%%%% BODY OF PAPER %%%%%%%%%%%%%%%%%%

\section{Introduction}
\label{sec:intro}

 The third-generation ground-based gravitational-wave (GW) interferometers, the Einstein Telescope \citep{punturo2010} and Cosmic Explorer \citep{reitze2019},  will  capture binary black hole (BBH) mergers up to a redshift $z\sim{100}$ \citep{maggiore2020,ng2021,ng2022}, with a factor of $\sim{100}$ higher sensitivity at $10$ Hz with respect to current detectors  \citep{maggiore2020, kalogera2022}. Hence, they will be the ideal observatories to probe the merger of stellar-sized black holes (BHs) in the early Universe \citep[e.g.,][]{ng2021,singh2022,ng2022b}, such as primordial BHs, and BHs born from Population~III  (hereafter, Pop.~III) stars. %stellar black holes (BHs), and primordial BHs \citep[e.g.,][]{ng2021,singh2022}. 

Here, we will focus on BHs born from the collapse of Pop.~III stars, i.e. the first, metal-free stars \citep{Haiman1996,Tegmark1997,Yoshida2003}. While we have not directly observed them yet, we expect that Pop. III stars gave a key contribution  to the reionization of the  Universe \citep{Kitayama2004,alvarez2006,Johnson2007} and to the enrichment of the intergalactic medium,  by spreading metals heavier than He though supernova explosions \citep[e.g.,][]{madau2001,bromm2003,tornatore2007,Karlsson2008,bromm2009,Karlsson2013}. 

%{\filippo{The initial mass function (IMF) of Pop. III stars is believed to be more top-heavy compared to that of metal-rich stars. This is mainly due to the inefficiency of molecular hydrogen as a coolant with respect to dust \citep[e.g.,][]{bromm2004,schneider2006,stacy2013,susa2014,hirano2014, hirano2015,wollenberg2020,chon2021,tanikawa2021,jaura2022,prole2022}. Massive Pop. III stars are known to retain nearly all of their mass throughout their life, as stellar winds are highly ineffective for a nearly metal-free chemical composition \citep[e.g.,][]{woosley2002,volpato2022}. If Pop. III stars can avoid pair instability \citep{woosley2017}, they may eventually experience a direct collapse, resulting in the formation of massive black holes  \citep[BHs, e.g.,][]{woosley2002}.}}{\filippo{Hence, }}

Mergers of BHs from Pop.~III stars have attracted a considerable interest \citep[e.g.,][]{ kinugawa2014,kinugawa2016,hartwig2016,belczynski2017,tanikawa2022} since the first LIGO--Virgo detection of a BBH merger, GW150914, with a total mass of $65.3^{+4.1}_{-3.4}$ M$_\odot$ in the source frame  \citep{abbottGW150914,abbottastrophysics}. In fact, BHs from Pop.~III stars are expected to extend to 
 higher masses than the compact remnants of Population~I  stars  (hereafter, Pop.~I stars, i.e.  metal-rich stars like our Sun) because mass loss by stellar winds is drastically quenched in  metal-free stars  \citep[e.g.,][]{madau2001,heger2002,woosley2002,schneider2002,kinugawa2014,Volpato2022}.  Moreover, the initial mass function of Pop.~III stars is commonly believed to be more top heavy than that of Pop.~I stars \citep[e.g.,][]{abel2002,bromm2004,Schneider2006, yoshida2006,stacy2013,bromm2013,glover2013,susa2014,hirano2014,hirano2015,wollenberg2020,chon2021, tanikawa2021, jaura2022, prole2022,klessen2023}, increasing the efficiency of BH formation. Also, Pop.~III binary stars tend to produce massive BBHs  because they are more likely to experience stable mass transfer than Pop.~I binary stars. In fact, massive Pop.~III stars tend to have  radiative envelopes for most of their life, avoiding common-envelope episodes \citep{kinugawa2016, inayoshi2017}.

For the above reasons, Pop.~III stars are among the main suspects for the formation of BHs inside or above the  pair-instability mass gap \citep[e.g.,][]{liubromm2020GW190521,farrell2021,kinugawa2021GW190521,tanikawa2021,tanikawa2022}, possibly explaining the formation of the peculiar merger GW190521, with primary (secondary) BH mass $85^{+21}_{-14}$ ($66^{+17}_{-18}$) M$_\odot$ \citep{abbottGW190521,abbottGW190521astro}.

Despite this revived interest in Pop.~III stars and their remnants, the actual merger rate density and mass spectrum of Pop.~III BHs are still debated \citep[e.g.,][]{kinugawa2016,belczynski2017,kinugawa2020}, mostly because of the absence of direct evidence for Pop.~III stars. Current predictions yield a local merger rate density of Pop.~III BBHs ranging from  $\sim{10^{-1}}$ to $\sim{10^2}$ yr$^{-1}$~Gpc$^{-3}$ \citep{kinugawa2014,belczynski2017,liubromm2020,tanikawa2022}.  This uncertainty comes from different assumptions regarding the initial binary properties, star and binary evolution processes, and star formation rate history. Moreover, dynamical interactions of Pop.~III BHs might also contribute to the merger rate \citep[e.g.,][]{liubromm2020,wang2022}. All of these uncertainties  propagate  into the redshift evolution of the mass spectrum and merger rate.

Here, we quantify the current uncertainties on the merger rate density and mass spectrum of BBH mergers from Pop.~III stars by considering a wide range of assumptions for the star formation history of metal-free stars \citep{jaacks2019,liu_bromm_20,skinner2020,hartwig2022}, for their initial binary properties \citep[e.g.,][]{larson1998,stacy2013,stacy2016,tanikawa2022}, and binary evolution \citep[e.g.,][and references therein]{costa2023}, by adopting the \sevn{} binary population synthesis code \citep{spera2019,mapelli2020,iorio22}.

\section{Methods}

\subsection{Population synthesis with {\sc{SEVN}}}
\label{sec:sevn}

We derived our BBH merger catalogues with the binary population synthesis code {\sc sevn}, which  integrates single and binary evolution by interpolating a set of pre-computed single stellar-evolution tracks, as described in \cite{iorio22}. Here, we adopt the following set-up of {\sc sevn}. We calculated the Pop.~III stellar tracks  with the {\sc parsec} code \citep{bressan2012,costa2021,nguyen2022} at metallicity $Z = 10^{-11}$. Here and in the rest of the manuscript, $Z$ is the mass fraction of elements heavier than helium, in absolute units.   This value of $Z$ is  equivalent to considering a metal-free composition \citep[e.g.,][]{marigo2001,tanikawa2021}. The zero-age main sequence mass (ZAMS) of our tracks ranges from 2 to 600~M$_\odot$. %\textcolor{red}{\textbf{
Tracks with $2<M_{\rm ZAMS}/{\rm M}_\odot< 8$ evolve until the end of the core He burning  and reach the early asymptotic giant branch phase, whereas tracks with  $M_{\rm ZAMS}> 8$ M$_\odot$ evolve until the beginning of the core O burning phase. Our tracks do not include stellar rotation and are computed with the same physical set-up as described by \citet{costa2021} for stellar winds, nuclear reaction network, opacity and equation of state. Above the convective core, we adopt a penetrative overshooting with a characteristic parameter of $\Lambda_\mathrm{ov} = 0.5$ in units of pressure scale height. We refer to \cite{costa2023} for more details on the evolutionary tracks.  In Section~\ref{sec:discussion}, we  compare the main features of our models with alternative Pop.~III models  \citep{tanikawa2021b,tanikawa2022b}. 
 We explore the uncertainties connected with different sets of stellar evolution models in a forthcoming work. %}}

%\textcolor{red}{\textbf{
Even if our initial stellar models are non spinning, \sevn{} includes a formalism for spin up and down via tides and mass accretion, based on  \cite{hurley2002}. However, we %do not have 
did not incorporate an accurate treatment for angular momentum transport in the stellar interior  \citep{Talon1997, Maeder1998, spruit2002} and we do not account for other processes that can affect the spin of the newly born compact object  during core collapse (e.g., the onset of an accretion disc). Hence, we will not discuss BH spin magnitudes here. %}}

We remap the final properties of the stars (in particular, final total mass  and CO core mass) into BH masses by adopting the rapid model for core-collapse supernovae \citep{fryer2012}. Furthermore, we implement the outcome of electron-capture supernovae, as detailed in \cite{giacobbo2019}. For (pulsational) pair-instability supernovae, we adopt the model presented in \cite{mapelli2020}.  In this model, based on the hydro-dynamical calculation by \cite{woosley2017} (see also \citealt{spera2017}), a star undergoes pulsational pair instability if the pre-supernova He-core mass, $M_\mathrm{He}$, is between 32 and 64 M$_{\odot}$. The mass of the BH after pulsational pair instability is:
\begin{equation}
    M_\mathrm{BH}=
    \begin{cases}
    
    \alpha_\mathrm{P} \,{}M_\mathrm{CCSN} & \mathrm{if}\,{}(\alpha_\mathrm{P} \,{}M_\mathrm{CCSN})\geq{}4.5\,{}\mathrm{M}_\odot\\
    0 & \mathrm{if}\,{}(\alpha_\mathrm{P} \,{}M_\mathrm{CCSN})<4.5\,{}\mathrm{M}_\odot,
    \end{cases}
\label{eq:PPImass}
\end{equation}
where $M_\mathrm{CCSN}$ is the mass of the BH after a core-collapse supernova (without pulsational pair instability) and $\alpha{}_{\rm P}$ is a dimensionless correction factor between 0 and 1. The dimensionless factor $\alpha_\mathrm{P}$ depends on $M_\mathrm{He}$ and the pre-supernova mass ratio between the mass of the He core and the total stellar mass  (see Equations 4 and 5 in the Appendix~of \citealt{mapelli2020}). For M$_\mathrm{He}> 64$ M$_\odot$, the star enters the pair-instability regime, and we assume that (i) the star is completely disrupted and leaves no compact remnant if M$_\mathrm{He}\leq{} 135$ M$_\odot$; (ii) the star directly collapses to a BH %for a larger final He mass 
if M$_\mathrm{He}> 135$ M$_\odot$ (see \citealt{costa2023} and \citealt{iorio22} for more details on these assumptions).

Here, we draw BH natal kicks from two different distributions. In our fiducial case, we adopt the formalism by \citet[][hereafter  \citetalias{giacobbo2020}]{giacobbo2020}:
\begin{equation}
V_{\rm kick}=f_{\rm H05}\,{}\frac{\langle{}M_{\rm NS}\rangle{}}{M_{\rm rem}}\,{}\frac{M_{\rm ej}}{\langle{}M_{\rm ej}\rangle},
\end{equation}    
 where $\langle{}M_{\rm NS}\rangle{}$ and $\langle{}M_{\rm ej}\rangle$ are the average neutron star mass and ejecta mass from single stellar evolution, respectively, while $M_{\rm rem}$ and $M_{\rm ej}$ are the compact object mass and the ejecta mass. The term $f_{\rm H05}$ is a random number drawn from a Maxwellian distribution with  one-dimensional root mean square $\sigma_\mathrm{kick}=265 \ \mathrm{km}\,{} \mathrm{s}^{-1}$, coming from a fit to the proper motions of 73 young pulsars ($<3$ Myr) in the Milky Way \citep{hobbs2005}.   In this formalism, stripped  and ultra-stripped  supernovae  result in lower kicks with respect to the other explosions, owing to the lower amount of ejected mass $M_{\rm ej}$ \citep{bray2016,bray2018}. BHs originating from a direct collapse receive zero natal kicks. 

In the alternative model we present in Section~\ref{sec:discussion},  we randomly draw the BH natal kicks from a Maxwellian distribution with one dimensional root mean square $\sigma_{\rm kick}=150$ km~s$^{-1}$ (hereafter $\sigma{}150$). This model matches the BH kicks inferred by \cite{atri2019}, based on the proper motions of 16 BH X-ray binaries in the Milky Way. The two models \citetalias{giacobbo2020} and $\sigma{}150$ bracket the uncertainties on BH natal kicks. While the latter is independent of the mass of the BH, the former introduces a strong dependence on both the mass of the compact remnant ($M_{\rm rem}$) and the evolution of the progenitor star (encoded in $M_{\rm ej}$).

In addition to the natal kick, we also calculate a Blaauw kick \citep{blaauw1961} resulting from the instantaneous mass loss in a binary system triggered by a supernova explosion. We use the same formalism as described in Appendix~A of \cite{hurley2002}.

Finally, {\sc sevn} integrates the following binary evolution processes: wind mass transfer, stable Roche-lobe overflow, common envelope evolution (adopting the $\alpha$ formalism, \citealt{hurley2002}), tidal evolution, stellar collisions, magnetic braking, and GW decay \citep{iorio22}.

Here, we use the same set-up as the fiducial model of \cite{iorio22}, adopting the default values for all relevant parameters (Section~3.2  of \citealt{iorio22}):
mass transfer is always stable for main sequence and Hertzsprung-gap donor stars, while we follow the 
prescriptions by \cite{hurley2002} in all the other cases. We set the Roche-lobe overflow mass accretion efficiency to 0.5 for a non-degenerate accretor, and assume that the mass which is not accreted  is lost from the vicinity of the accretor as an isotropic wind (isotropic re-emission). 
At the onset of the Roche-lobe overflow, {\sc sevn} circularises the orbit at  periastron.  During common envelope, we estimate the envelope binding
energy using the same  formalism as in \cite{claeys2014}.  We adopt $\alpha{}=1$ for the common-envelope efficiency parameter, i.e. we assume that all the kinetic energy lost from the system contributes to unbinding the common envelope. %\textcolor{red}{\textbf{
In Appendix~\ref{app:mass_transfer}, we discuss the impact of different assumptions for mass accretion and common-envelope efficiency. %}}

\subsection{Initial conditions for Pop.~III binary systems}
\label{sec:IC}

%For Pop.~III binary stars, 
We use the same binary-population synthesis simulations as in \cite{costa2023}. 
We summarise their initial conditions here below and in Table~\ref{tab:IC}.

\subsubsection{Initial mass function (IMF)}
\label{sec:IMF}

Among the many different models proposed in the literature, we consider the following four distributions for the initial mass function (IMF), %in order to account for 
because they bracket the uncertainties on the IMF of Pop.~III stars \citep[e.g.,][]{bromm2004,yoshida2006,bromm2013, glover2013}. 

\begin{itemize}
    \item A flat-in-log probability distribution function  $\xi(M_{\rm ZAMS})$
    \citep[see e.g.][]{stacy2013,susa2014,hirano2014, hirano2015,wollenberg2020,chon2021,tanikawa2021,jaura2022,prole2022}:
    %(hereafter, LOG, \citealt{stacy2013,susa2014,hirano2014, hirano2015,wollenberg2020,chon2021,tanikawa2021,jaura2022,prole2022}: %in agreement with %\cite{susa2014}, \cite{hirano2014, hirano2015}, and \cite{tanikawa2021,tanikawa2022}:  
    \begin{equation}
        \xi(M_{\rm ZAMS}) \propto M_{\rm ZAMS}^{-1}.
    	\label{eq:IMF_Logflat}
    \end{equation}
    
    \item A \citet{kroupa2001} distribution (hereafter \citetalias{kroupa2001}):
    \begin{equation}
        \xi(M_{\rm ZAMS}) \propto M_{\rm ZAMS}^{-2.3}.
    	\label{eq:IMF_Kroupa}
    \end{equation}
    %Our distribution does not change slope at low mass, because we consider only  masses $\geq{}2.2$~M$_\odot$.
With respect to the original \citetalias{kroupa2001}, which has a flatter slope for $M_{\rm ZAMS}<0.5$ M$_\odot$, here we assume a single slope because we do not generate ZAMS masses $<5$ M$_\odot$ from this distribution. 
    
    \item A \citet{larson1998} distribution (hereafter \citetalias{larson1998}):
    \begin{equation}
        \xi(M_{\rm ZAMS}) \propto M_{\rm ZAMS}^{-2.35} e^{- M_\mathrm{cut1}/M_{\rm ZAMS}},
    	\label{eq:IMF_Larson}
    \end{equation}
    where $M_{\rm cut1} = 20$ M$_\odot$ \citep{valiante2016}.
    
    \item A top-heavy distribution (hereafter TOP), following \citet{stacy2013}, \citet{jaacks2019}, and \citet{liubromm2020}:
    \begin{equation}
        \xi(M_{\rm ZAMS}) \propto M_{\rm ZAMS}^{-0.17} e^{- M_\mathrm{cut2}^2/M_{\rm ZAMS}^2},
    	\label{eq:IMF_Topheavy}
    \end{equation}
    where $M_\mathrm{cut2} = 20$~M$_\odot$.
\end{itemize}

 In the following, we call LOG, KRO, LAR, and TOP our models adopting the flat-in-log, \citetalias{kroupa2001}, \citetalias{larson1998}, and top-heavy IMFs, respectively (Table~\ref{tab:IC}). In all of our models but LOG3 (Table~\ref{tab:IC}), we use the aforementioned IMFs to generate the ZAMS mass of the primary star $M_{\rm ZAMS,1}$ (i.e., the most massive component of the binary star) in the range $[5,550]$~M$_\odot$. In model LOG3, we instead randomly sample the entire IMF (both primary and secondary stars) in the range  $M_{\rm ZAMS}\in{[5,\,{}550]}$~M$_\odot$ according to the LOG IMF.

\subsubsection{Mass ratio and secondary mass}
\label{sec:q}

We draw the mass of the secondary star ($M_{\rm ZAMS,\,{}2}$)  according to three different distributions. 

\begin{itemize}
    \item We use the distribution of the mass ratio $q=M_{\rm ZAMS,2}/M_{\rm ZAMS,1}$ from  \citet[][hereafter  \citetalias{sana2012}]{sana2012}:
    \begin{equation}
        \xi(q) \propto q^{-0.1} \; {\rm with} \; q \in [0.1, 1] \; {\rm and} \; M_{\rm ZAMS,2} \geq 2.2~{\rm M}_{\odot}.
	    \label{eq:Sana}
    \end{equation}
     This distribution is a fit to the mass ratio of O- and B-type binary stars in the local Universe.% \citep{sana2012}.   
    
    \item In the sorted  distribution, we draw the ZAMS mass of the entire stellar population from the same IMF, and then we randomly pair two stars from this distribution, imposing that $M_{\rm ZAMS,2}\leq{}M_{\rm ZAMS,1}$. 
    In this model,  the  minimum mass of the secondary is equal to that of  the primary (5~M$_\odot$) by construction. 
    
    \item The mass ratio distribution by \citet[][hereafter \citetalias{stacy2013}]{stacy2013}: 
    \begin{equation}
        \xi(q) \propto{} q^{-0.55}\; {\rm with} \; q \in [0.1, 1] \; {\rm and} \; M_{\rm ZAMS,2} \geq 2.2~{\rm M}_\odot.
    	\label{eq:q_stacy}
    \end{equation}
    This distribution was obtained from a fit to Pop.~III stars formed in %mini-halos of 
    cosmological simulations (\citetalias{stacy2013}).

\end{itemize}

\subsubsection{Orbital period}
\label{sec:P}
We consider two different distributions for the orbital period ($P$):
\begin{itemize}
    \item The distribution derived by \citetalias{sana2012} for O- and B-stars in the local Universe:  %favours tight  binaries with a peak at $P\sim{10}$~days:
    \begin{equation}
        \xi(\pi) \propto{} \pi^{-0.55} \quad {\rm with} \quad \pi = \log (P/{\rm day}) \in [0.15, 5.5].
    	%\label{eq:P_sana}
    \end{equation}
    
    \item A Gaussian distribution
    \begin{equation}
        \xi(\pi) \propto{} \exp{\left[-(\pi-\mu)^2/(2\,{}\sigma{}^2)\right]}
    	%\label{eq:P_sana}
    \end{equation}
    with $\mu{}=5.5$, and  $\sigma{}=0.85$, as derived from \citetalias{stacy2013} based on cosmological simulations.  While this distribution is likely affected by the numerical resolution of the original simulations, which suppresses the formation of systems with short orbital periods, we decide to consider it as a robust upper limit to the orbital period of Pop.~III binary stars.
    %The period distribution found by \citetalias{stacy2013} is a Gaussian distribution of $\log P$ with a peak at $\sim 5.5$, and a standard deviation of  $\sigma \sim 2$, favouring long periods with respect to the S12 distribution.  
\end{itemize}

\subsubsection{Eccentricity}
\label{sec:e}

We compare two distributions for the orbital eccentricity $(e)$:
\begin{itemize}
    \item The distribution obtained by  \citetalias{sana2012} and based on a sample of O- and B-type stars in the local Universe:
    \begin{equation}
        \xi(e) \propto e^{-0.42} \quad {\rm with} \; e \in [0, 1).
    	\label{eq:ecc_sana}
    \end{equation}

    \item The thermal distribution, adopted  for Pop.~III binaries by, e.g., \cite{kinugawa2014,hartwig2016,tanikawa2021}:
    \begin{equation}
        \xi(e) =2\,{} e\quad {\rm with} \; e\in [0, 1).
    	\label{eq:ecc_therm}
    \end{equation}
\end{itemize}

\subsubsection{Input catalogues}
\label{sec:ICc}

We build  11 different input catalogues by varying the aforementioned distributions  of 
the IMF, $q$, $P$, and $e$. We set the total number of generated binaries to  obtain  $10^7$ binaries in the high-mass regime ($M_{\rm ZAMS,2} \geq 10~{\rm M}_\odot$, and $M_{\rm ZAMS,1} \geq 10~{\rm M}_\odot$ by construction).  This %sampling strategy 
ensures a good sampling of the high-mass regime and reduces the stochastic fluctuations  \citep[e.g.,][]{iorio22}. Table~\ref{tab:IC} lists the properties of our input catalogues. We refer to \cite{costa2023} for more details.

%%%%%%%%%%%%%%%%%%%TABLE%%%%%%%%%%%%%%%%%%%%%%%
\begin{table}
    \caption{Initial conditions.}  
    \begin{center}
        \begin{tabular}{cccccc}
            \hline
            Model &  $M_{\rm ZAMS,1}$  & $M_{\rm ZAMS}$   &  $q$     & $P$      &  $e$ \\ %&  $N$ [$\times{}10^7$] & Total mass [$\times{}10^9$~\Msun] \\ 
            \hline
            LOG1     & Flat in log  & -- & \citetalias{sana2012}         & \citetalias{sana2012}       & \citetalias{sana2012}\\%    & 1.45 & 2.59       \\
            LOG2     & Flat in log  & --& \citetalias{sana2012}         & \citetalias{stacy2013}           & Thermal  \\ %& 1.45 &  2.58    \\
            LOG3     & --  & Flat in log & Sorted         & \citetalias{sana2012}         & \citetalias{sana2012}   \\ %& 1.38 &  3.19    \\
             LOG4   & Flat in log  & -- & \citetalias{stacy2013}           & \citetalias{sana2012}         &   Thermal  \\ %& 1.53 & 2.60      \\
            LOG5     & Flat in log & -- & \citetalias{stacy2013}           & \citetalias{stacy2013}           &  Thermal  \\ %& 1.53 & 2.60 \\
            \hline
            KRO1     & \citetalias{kroupa2001}  & -- & \citetalias{sana2012}         & \citetalias{sana2012}         & \citetalias{sana2012}  \\ % & 5.23 (2.00$\dagger$)  & 1.35 (0.89$\dagger$)      \\
            KRO5     & \citetalias{kroupa2001}   & -- & \citetalias{stacy2013}           & \citetalias{stacy2013}           & Thermal  \\ %& 6.11 (2.00$\dagger$)   & 1.52 (0.93$\dagger$)  \\ 
            \hline
            LAR1     & \citetalias{larson1998}  & -- & \citetalias{sana2012}         & \citetalias{sana2012}          & \citetalias{sana2012}  \\% & 2.00    & 1.20     \\
            LAR5     & \citetalias{larson1998}  & -- &  \citetalias{stacy2013}           & \citetalias{stacy2013}            & Thermal  \\ %&  2.27 (2.00$\dagger$)  & 1.30 (1.24$\dagger$) \\ 
            \hline
            TOP1     & Top heavy & -- & \citetalias{sana2012}        & \citetalias{sana2012}         & \citetalias{sana2012}   \\ % & 1.05 &  4.16     \\
            TOP5     & Top heavy & -- & \citetalias{stacy2013}          & \citetalias{stacy2013}           & Thermal  \\ % &  1.07 & 4.03  \\ 
            \hline
        \end{tabular}
    \end{center}
    \footnotesize{Column~1 reports the model name. Column~2 describes how we generate the ZAMS mass of the primary star (i.e., the most massive of the two members of the binary system). Column~3 describes how we generate the ZAMS mass of the overall stellar population (without differentiating between primary and secondary stars). We follow this procedure only for model LOG3 (see the text for details).
    Columns~4, 5, and 6 specify the distributions we used to generate the mass ratios $q$, the orbital periods $P$ and the orbital eccentricity $e$.
    See Section~\ref{sec:IC} for a detailed description of these distributions.}   
    \label{tab:IC} 
\end{table}
%%%%%

\subsection{Formation history of Pop.~III stars }
\label{sec:sfrd}

\begin{figure}
    \centering
    \includegraphics[width =  \columnwidth]{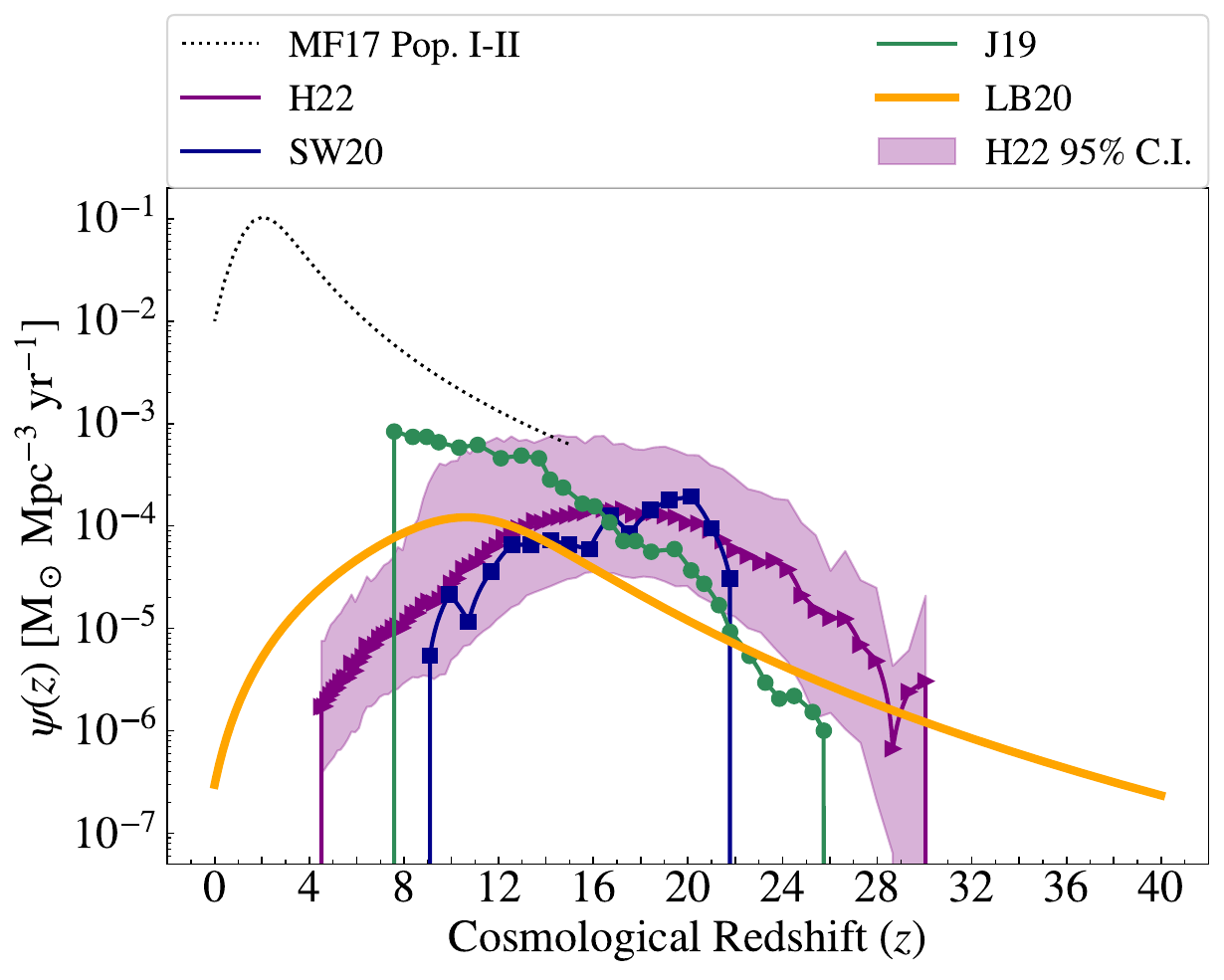}
    \caption{Star formation rate density $\psi(z)$   of Pop. III stars. Purple line: 
 \protect\cite{hartwig2022} (\citetalias{hartwig2022}); blue line: \protect\cite{skinner2020} (\citetalias{skinner2020}); green line: \protect\cite{jaacks2019} (\citetalias{jaacks2019}); orange line \protect\cite{liubromm2020} (\citetalias{liu_bromm_20}).  
 The shaded area shows the 95\% credible interval for the \citetalias{hartwig2022} model. 
    The thin dotted black line is the star formation rate density of  Pop. I-II stars from \protect \cite{madau2017} (MF17).}
    \label{fig:sfrd}
\end{figure}

We consider four independent estimates of the Pop.~III star formation rate density (SFRD), to quantify the main differences among models. Figure \ref{fig:sfrd} shows these four star formation rate histories, which come from \citet[][hereafter \citetalias{hartwig2022}]{hartwig2022},  \citet[][hereafter 
 \citetalias{jaacks2019}]{jaacks2019},  \citet[][hereafter \citetalias{liu_bromm_20}]{liu_bromm_20}, and \citet[][hereafter \citetalias{skinner2020}]{skinner2020}. All of them are consistent with the value of the Thomson scattering optical depth estimated by the Planck Collaboration \citep[i.e. $\tau_e = 0.0544 \pm 0.0073$, ][]{Planck2016}.

 The peak of the Pop.~III SFRD is remarkably different in these four models, ranging from $z\approx{8}$ (\citetalias{jaacks2019}) to $z\approx{20}$ (\citetalias{skinner2020}). By selecting these four SFRDs, we account for  differences in the underlying physics assumptions and for  the cosmic variance, since  these models rely on cosmological boxes with length spanning from 1 to 8 $h^{-1}$ comoving Mpc. %Hereafter, we briefly describe the setup of the SFRD models we considered. 
All of these SFRDs come from semi-analytic  models or cosmological simulations that assume their own IMF for Pop.~III stars. Thus, we introduce an inconsistency whenever we combine a given SFRD model with a binary population synthesis catalogue generated with a different IMF. We expect that the impact of this assumption on our results is negligible  compared to other sources of uncertainty considered in this work \citep[e.g.,][]{crosby2013}. 

%%%%%%%%%%%%FIGURE%%%%%%%%%%%%%%%%%%%%%%%
%%%%%%%%%%%%%%%%%%%%%%%%%%%%%%

\subsubsection{\citetalias{hartwig2022}}
\label{sec:asloth}

\asloth is the only semi-analytic model that samples and traces individual Pop. III and II stars, and is calibrated on several observables from the local and high-redshift Universe \citep{hartwig2022,uysal2023}. It can take  input dark-matter halo merger trees either from cosmological simulations or from an extended Press-Schechter formalism. Here, we use the results obtained with the merger tree from  \cite{ishiyama2016}, who simulated a (8 $h^{-1}$ Mpc)$^3$ box down to a redshift $z=4$. 

With {\sc{a-sloth}}, it is possible to quantify the uncertainties in the SFRD that arise from unconstrained input parameters of the semi-analytic model (such as star formation efficiency or Pop.~III IMF). Hartwig et al., in prep.,  characterise these uncertainties through a Monte-Carlo Markov Chain (MCMC) exploration. They calibrate {\sc{a-sloth}} with a rejection sampler that should maximise the log likelihood, which is based on nine independent observables. After the initial burn-in phase, they record every accepted model. In this way, they explore the parameter space around the optimum and obtain various different models that all reproduce observables equally well. The MCMC runs provide $>5000$ accepted models, each with a slightly different SFRD. From these models, we have extracted the central 95\% credible interval of the SFRD. Figure \ref{fig:sfrd} shows this 95\% credible interval, which reflects uncertainties in the unconstrained input parameters.

\subsubsection{\citetalias{jaacks2019}}

\cite{jaacks2019} used the hydro-dynamical/$N$-body code  {\sc gizmo} \citep{gizmo} coupled with  custom sub-grid physics, % to reproduce Pop. III star formation, in particular 
accounting for both the chemical and radiative feedback from core-collapse and pair-instability supernovae. The simulation has been run down to  $z = 7.5$  with a comoving box length of 4 $h^{-1}$ Mpc.

\subsubsection{\citetalias{liu_bromm_20}}

\cite{liu_bromm_20} also ran a cosmological simulation with {\sc{gizmo}}, but assumed different sub-grid prescriptions,
resulting in a lower Pop.~III SFRD  compared to \cite{jaacks2019}. They simulated a comoving cubic box of 
(4 $h^{-1}$ Mpc)$^3$ down to redshift $z=4$, and then  extrapolate the results to $z=0$ with additional semi-analytic modelling. 
They parameterised their Pop.~III SFRD evolution with the same functional form as \cite{madau2014} 
\begin{equation}\label{eq:liu}
        \psi(z) = \frac{a\,{}(1+z)^b}{1+[(1+z)/c]^d}~~[{\rm{M}}_\odot~{\rm{yr}}^{-1}~ {\rm{Mpc}}^{-3}],
\end{equation}
and obtained best-fit parameters  $a = 756.7~{\rm{M}}_\odot~{\rm{yr}}^{-1}~ {\rm{Mpc}}^{-3}$, $b=-5.92,~c= 12.83$, and $d=-8.55$ \citep{liu_bromm_20}. In our analysis, we use this best fit.

\subsubsection{\citetalias{skinner2020}}

\cite{skinner2020} ran a hydro-dynamical cosmological simulation with the adaptive mesh refinement code {\sc{Enzo}} \citep{Bryan2014}. They simulate a (1 $h^{-1}$ Mpc)$^{3}$ comoving box with a 256$^{3}$ base grid resolution and a dark-matter particle mass of 2001 M$_\odot$. %They applied a time-dependent Lyman-Werner (LW) optically thin radiation background. 
This simulation has been run down to $z = 9.32$.

\subsection{{\sc{cosmo$\mathcal{R}$ate}}}
\label{sec:cosmorate}

%%%%%%%%%%%%%%%%%%%%FIGURE%%%%%%%%%%%%%%%%%%%%
\begin{figure}
    \centering
    \includegraphics[width =  \columnwidth]{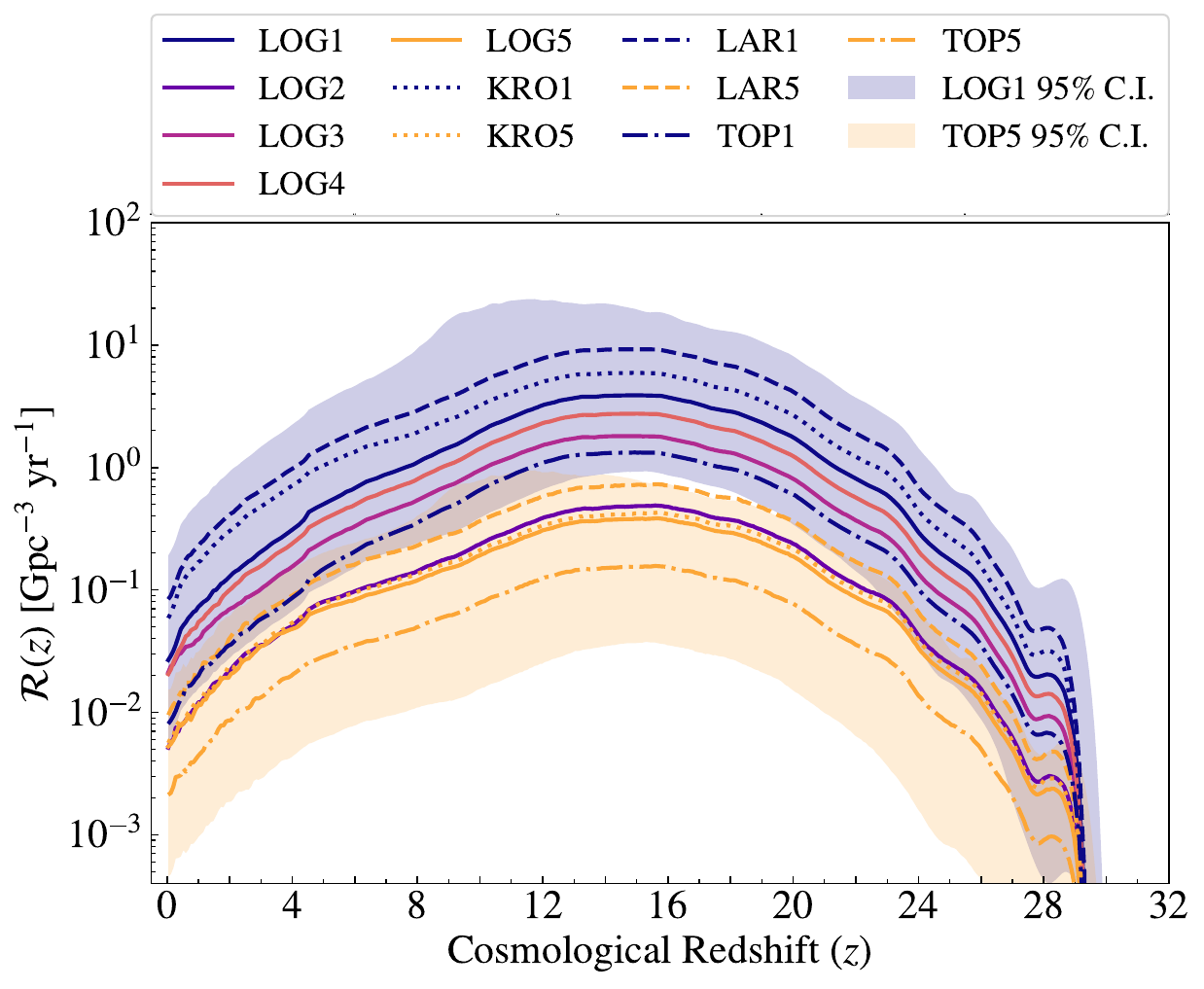}
    \caption{Evolution of the BBH merger rate density with redshift $\mathcal{R}(z)$ assuming the SFRD from \citetalias{hartwig2022}, and the corresponding 95\% credible interval. Solid lines: models with a flat-in-log IMF (LOG). Dotted lines: Kroupa IMF (KRO). Dashed lines: Larson IMF (LAR). Dot-dashed lines: top-heavy IMF (TOP). The shaded areas are 95\% credible interval evaluated considering input uncertainty in {\sc a-sloth} (see Figure \protect\ref{fig:sfrd} and Section \ref{sec:asloth}) for the models LOG1 and TOP5.}
    \label{fig:mrd_asloth_uncrt}
\end{figure}
%%%%%%%%%%%%%%%%%%%%%%%%%%%%%%%%%%%%%%%%%%

%%%%%%%%%%%%FIGURE%%%%%%%%%%%%%%%%%%%%%%%%%
\begin{figure*}
    \centering

    \includegraphics[width = 0.9 \textwidth]{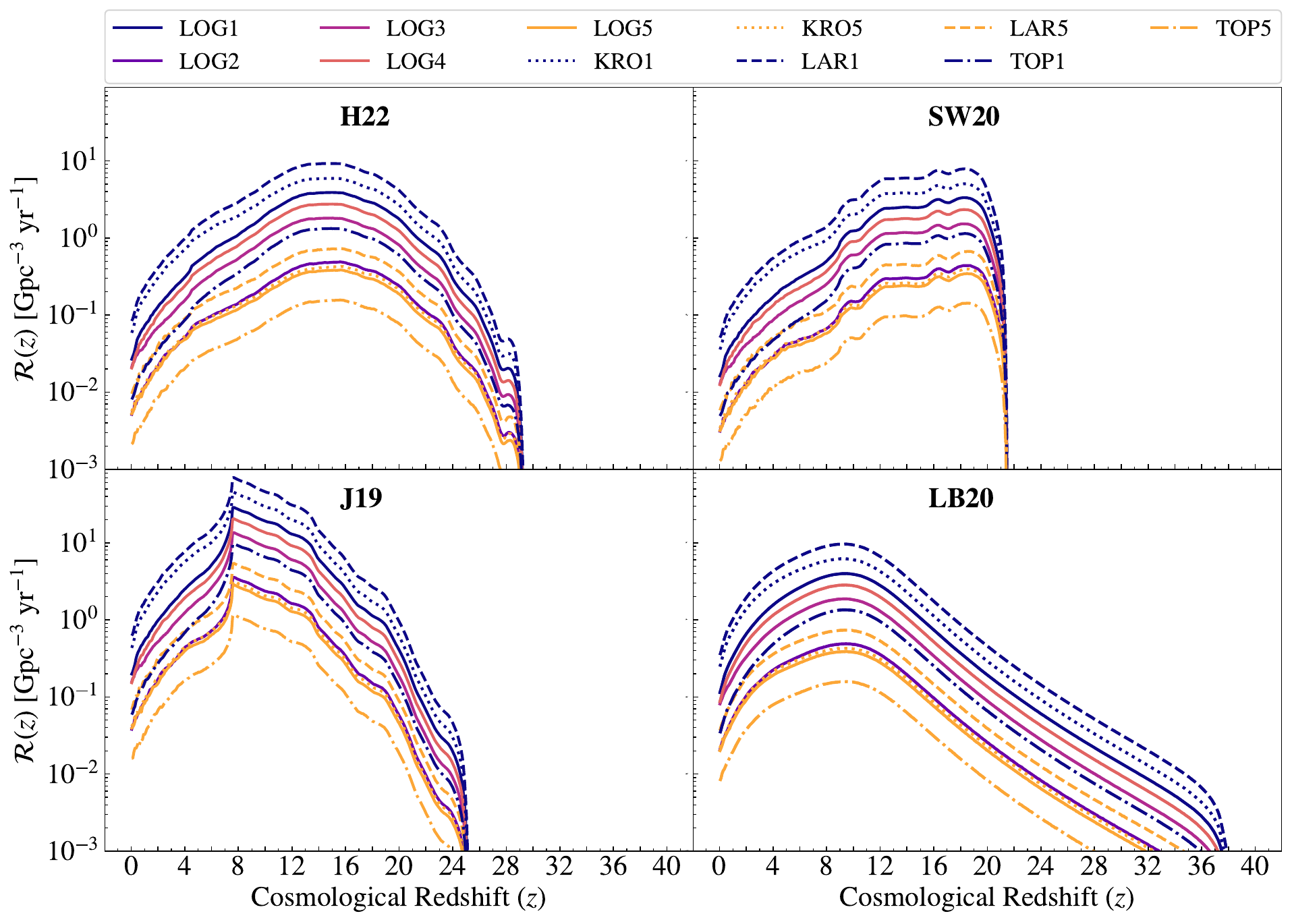}
    \caption{Evolution of the BBH merger rate density with redshift $\mathcal{R}(z)$ for all the 44 models considered in this work. Solid lines: models with a flat-in-log IMF (LOG). Dotted lines: Kroupa IMF (KRO). Dashed lines: Larson IMF (LAR). Dot-dashed lines: top-heavy IMF (TOP).  Upper left plot: \citetalias{hartwig2022} star formation history, upper right: \citetalias{skinner2020}, lower left: \citetalias{jaacks2019}, and lower right: \citetalias{liu_bromm_20}.} 
    \label{fig:mrd_pro}
\end{figure*}
%%%%%%%%%%%%%%%%%%%%%%%%%%%%%%%%%%%%%%%%%%

We estimate the merger rate density evolution of BBHs with the semi-analytic code \cosmorate{} \citep{santoliquido2020,santoliquido2021}, which interfaces catalogues of 
simulated BBH mergers with a metallicity-dependent SFRD  model. %Since we model Pop.~III stars with a single metallicity ($Z=10^{-11}$), t
The merger rate density %of Pop.~III BHs 
in the comoving frame  is given by
\begin{equation}
\label{eq:mrd}
    \mathcal{R}(z) = \int_{z_{{\rm{max}}}}^{z}\left[\int_{Z_{{\rm{min}}}}^{Z_{{\rm{max}}}} \,{}\mathcal{S}(z',Z)\,{} %\eta(Z)
    \mathcal{F}(z',z,Z) \,{}{\rm{d}}Z\right]\,{} \frac{{{\rm d}t(z')}}{{\rm{d}}z'}\,{}{\rm{d}}z',
\end{equation}
where $\mathcal{S}(z',Z)= \psi(z')\,{}p(z',Z)$. %\]
%\end{equation}
%In Equation~\ref{eq:sfrd}, 
Here, $\psi(z')$ is the adopted  star formation rate density evolution (chosen among the ones presented in Figure \ref{fig:sfrd}), and $p(z',Z)$ is the distribution of metallicity $Z$ at fixed formation redshift $z'$. %\textcolor{red}{\textbf{
Since we model Pop.~III stars  with a single metallicity ($Z=10^{-11}$), we define  $p(z',Z)$ as a delta function for Pop.~III stars, different from zero only if $Z=10^{-11}$. %}}
In Equation \ref{eq:mrd}, ${{\rm{d}}t(z')}/{{\rm{d}}z'} =  H_{0}^{-1}\,{}(1+z')^{-1}\,{}[(1+z')^3\Omega_{M}+ \Omega_\Lambda{}]^{-1/2}$, where $H_0$ is the Hubble  parameter, $\Omega_M$ and $\Omega_\Lambda$ are the matter and energy density, respectively. We adopt the values  of \cite{Planck2018}. The term $\mathcal{F}(z',z, Z)$ in Equation~\ref{eq:mrd} is given by:
\begin{equation}
\mathcal{F}(z',z, Z) = \frac{1}{\mathcal{M}_{{\rm{TOT}}}(Z)}\frac{{\rm{d}}\mathcal{N}(z',z, Z)}{{\rm{d}}t(z)},
\end{equation}
where $\mathcal{M}_{{\rm{TOT}}}(Z)$ is the total simulated initial stellar mass in our binary population-synthesis simulations, and  ${{\rm{d}}\mathcal{N}(z',z, Z)/{\rm{d}}}t(z)$ is the rate of BBHs that form from progenitor metallicity $Z$ at redshift $z'$ and merge at $z$, extracted from our population-synthesis catalogues.

For all the Pop.~III models shown in this work, we assume that the binary fraction is $f_{\mathrm{bin}} = 1$, and we do not apply any correction for not sampling stars with mass  $< 5$~M$_\odot$. We make this simplifying assumption because we do not know the minimum mass and binary fraction of Pop.~III stars. Assuming a lower binary fraction and a lower minimum mass than  $m_{\rm min}=5$ M$_\odot$ translates into a shift of our merger rate by a constant numerical factor (unless we assume that the minimum mass and the binary fraction depend on either redshift or metallicity). 
%, i.e. $f_{\mathrm{IMF}} = 1$.}}

\subsection{Einstein Telescope detection rate}
\label{app:r_det_method}

 We evaluate the detection rate of Pop.~III BBH mergers ($\mathcal{R}_{\mathrm{det}}$) by the Einstein Telescope, as follows \citep[e.g.,][]{fishbach2021, broek2022}:
\begin{equation}
    \label{eq:rdet}
    \mathcal{R}_\mathrm{det} = \int 
    \frac{\mathrm{d}^2\mathcal{R}(m_1,m_2, z)}{\mathrm{d}m_1\mathrm{d}m_2}\frac{1}{(1+z)} \frac{\mathrm{d}V_c}{\mathrm{d}z} \,{}p_{\mathrm{det}}(m_1, m_2, z)~ \mathrm{d}m_1\,{}\mathrm{d}m_2\,{}\mathrm{d}z,
\end{equation}
where the distribution of merger rate density as a function of primary $m_1$ and secondary mass $m_2$ is given by
\begin{equation}
    \frac{\mathrm{d}^2\mathcal{R}(m_1,m_2,z)}{\mathrm{d}m_1\mathrm{d}m_2} = \mathcal{R}(z)\,{}p(m_1,m_2 |z).\label{eq:dR}
\end{equation}
In Equation \ref{eq:dR},  $\mathcal{R}(z)$ is the merger rate density as a function of redshift (in units of Gpc$^{-3}$ yr$^{-1}$, Equation \ref{eq:mrd}) and $p(m_1,m_2 |z)$ is the two-dimensional source-frame mass distribution at a given redshift extracted with {\sc{cosmo$\mathcal{R}$ate}} for each astrophysical model (Table \ref{tab:IC}). The factor $1/(1+z)$ converts source-frame time to detector-frame time and $\mathrm{d}V_c/\mathrm{d}z$ is the differential comoving volume element. 

In Equation \ref{eq:rdet}, $p_{\mathrm{det}}(m_1, m_2, z)$ is the probability of detecting a single system with parameters $m_1$, $m_2$, and $z$. We assume %the simplification 
that $p_{\mathrm{det}}(m_1, m_2, z)$ is an Heaviside step function, i.e. $p_{\mathrm{det}}(m_1, m_2, z) = 1$ only if $\rho > \rho_{\mathrm{th}}$, where $\rho$ is the total signal-to-noise ratio (SNR) and $\rho_{\mathrm{th}} = 9$. 
To evaluate the total SNR, we assume that the Einstein Telescope is composed of three independent, identical, and triangular-shaped detectors. Therefore, we evaluate the total SNR as \citep[e.g.,][]{singh2021,yi2022}:
\begin{equation}
\label{eq:tot}
    \rho = \rho_{\mathrm{opt}}\,{}\sqrt{\omega_0^2+\omega_1^2+\omega_2^2}
\end{equation}
where $\rho_\mathrm{opt}$ is the optimal SNR \citep{finn1993, dominik2015,taylor2018,bouffanais2019,chen2021}:
\begin{equation}
    \label{eq:snr_opt}
    \rho_{\mathrm{opt}}^2 = 4\int_{f_{\mathrm{low}}}^{f_{\mathrm{high}}} \mathrm{d}f~ \frac{|\tilde{h}(f)|^2}{S_n(f)}
\end{equation}
with $f_{\mathrm{low}} = 2$ Hz and $f_{\mathrm{high}} = 1000$ Hz.  The frequency domain response of ET ($\tilde{h}(f)$) to a face-on non-precessing BBH merger signal with $m_1$, $m_2$ and $z$ is generated using {\sc{pyCBC}} \citep{Biwer2019} and assuming {\sc{IMRPhenomXAS}} as the phenomenological waveform
model \citep{garciaquiros2020}. In Equation~\ref{eq:snr_opt}, 
 $\omega_i = \Theta_i/4$ where $i = 0, 1, 2$ is the index of the interferometer and $\Theta_i$ is the angular dependence of the GW signal, defined as:
\begin{equation}
    \Theta_i = 2[F_{+,i}^2(1+\cos^2\iota)^2+4F_{\times,i}^2\cos^2\iota]^{1/2}
\end{equation}
with $0 < \Theta_i < 4$ \citep{finn1993,singh2021} and $\cos \iota$ is the cosine of the  inclination angle randomly sampled in the range $[-1,1]$. The Einstein Telescope will have three nested interferometers, rotated by 60\degree{} with respect to each other \citep{regimbau2012}. The antenna pattern for each interferometer is thus $F_{+,i}(\theta, \phi, \psi) = F_{+,0}(\theta, \phi + 2i\pi/3, \psi)$ and $F_{\times,i}(\theta, \phi, \psi) = F_{\times,0}(\theta, \phi + 2i\pi/3, \psi)$, where $\theta$, $\phi$ denote the sky position and $\psi$ is the polarisation angle. The response function $F_{+,0}$ and $F_{\times,0}$ are given  as \citep{finn1993,regimbau2012,singh2021}:
\begin{equation}
    F_{+,0}  = \frac{\sqrt{3}}{4}(1+\cos^2 \theta)\cos2\phi \cos2\psi - \frac{\sqrt{3}}{2}\cos\theta \sin2\phi \sin2\psi
\end{equation}
and
\begin{equation}
    F_{\times,0} = \frac{\sqrt{3}}{4}(1 + \cos^2 \theta) \cos 2\phi \sin 2\psi + \frac{\sqrt{3}}{2}\cos \theta \sin 2\phi \cos 2\psi.
\end{equation}
We assume all the sources to be isotropically distributed, therefore we randomly sample $\cos \theta$, $\phi$, and $\psi$ in the ranges [-1,1], $[-\pi, \pi]$, and $[-\pi, \pi]$, respectively. In  Equation~\ref{eq:snr_opt}, $S_n(f)$ is the noise power spectral density and represents the sensitivity of the detector to GWs at different frequencies. In our calculation, we adopt the ET-D 10-km triangle  configuration for $S_n(f)$ \footnote{ET sensitivity curves are available at \href{https://www.et-gw.eu/index.php/etsensitivities}{https://www.et-gw.eu/index.php/etsensitivities}} \citep[][]{hild2008,hild2010,hild2011}.

\section{Results}

\subsection{Merger rate density of BBHs born from Pop.~III stars}
\label{sec:mrd}

Figure \ref{fig:mrd_asloth_uncrt} shows the merger rate density evolution of Pop. III BBHs %as a function of redshift evaluated with 
assuming the SFRD from \citetalias{hartwig2022}. The merger rate density changes by about one order of magnitude within the 95\% credible interval of the Pop.~III SFRD estimated by \citetalias{hartwig2022}. 
%This will have relevant implications  for the future science case of third-generation detectors. Moreover, the 
Uncertainties on the initial conditions of binary systems (Table \ref{tab:IC}) impact the merger rate density of Pop.~III BBHs by up to two orders of magnitude, for a fixed SFRD model. The models adopting a \citetalias{stacy2013} distribution for the initial orbital periods (LOG2, LOG5, KRO5, LAR5, and TOP5) have lower merger rate densities than models adopting the distribution by \citetalias{sana2012} (all the remaining models). The reason is that short  orbital periods, as in the case of \citetalias{sana2012}, favour the merger of BBHs via stable mass transfer episodes between the progenitor stars.

Figure \ref{fig:mrd_pro} shows the merger rate density for all the SFRD models considered in this work. The star formation rate history affects both the shape and the normalisation of the BBH merger rate density. Our merger rate density curves (Fig.~\ref{fig:mrd_pro}) are similar in shape to the SFRD curves (Fig.~\ref{fig:sfrd}), with just a shift to lower redshift because of the delay time, i.e. the time between the formation of a BBH-progenitor binary system and the merger of the two BHs.   %, because our binary evolution models have relatively short delay times. 
Hence, the peak of the merger rate density spans from $z\approx{16}$ to $z\approx{8}$ depending on the SFRD model. For all the considered star formation rate histories and binary models, the BBH merger rate density peaks  well inside the instrumental horizon of the Einstein Telescope \citep{maggiore2020,kalogera2022}.

\subsection{Detection rate of Pop.~III BBHs with the Einstein Telescope}

%%%%%%%%%%%%%%%%FIGURE%%%%%%%%%%%%%%%%%%%%%%
\begin{figure}
    \centering
    \includegraphics[width =  \columnwidth]{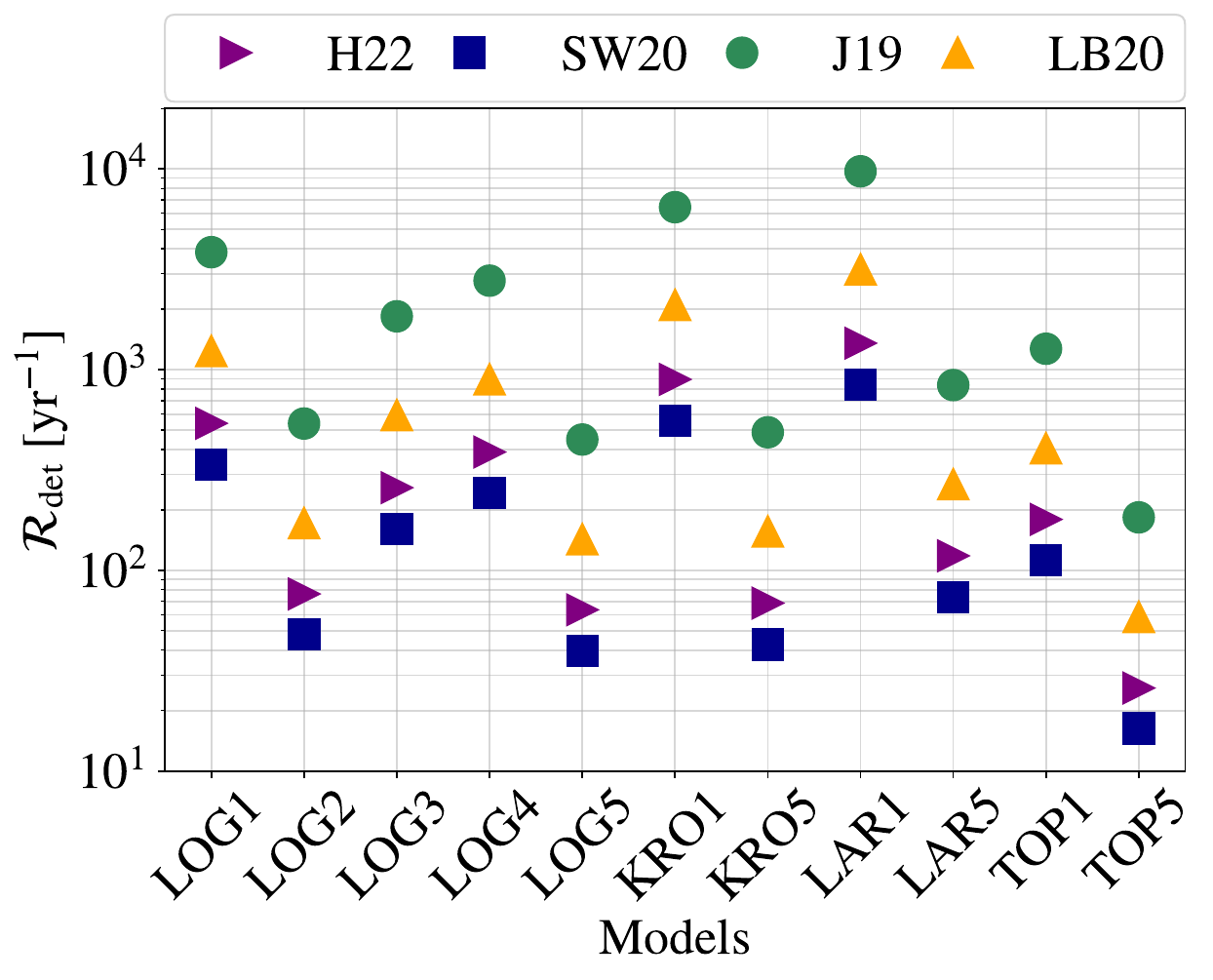}
    \caption{Detection rate $\mathcal{R}_{\mathrm{det}}$ of Pop.~III BBHs, assuming the Einstein Telescope triangle configuration (ET-D) and $\rho_{\mathrm{th}} = 9$. See Section~\ref{app:r_det_method} for details. Purple right-pointing triangles: SFRD from \citetalias{hartwig2022}; blue squares: \citetalias{skinner2020}; green circles: \citetalias{jaacks2019}; orange triangles: \citetalias{liu_bromm_20}. }
    \label{fig:r_det}
\end{figure}
%%%%%%%%%%%%%%%%%%%%%%%%%%%%%%%%%%%%%%%%%%%%

We have estimated the detection rate of Pop.~III BBH mergers that we expect to obtain with the Einstein Telescope, in the 10-km triangle configuration (ET-D, \citealt{hild2008,hild2010,hild2011}), as described in Section~\ref{app:r_det_method}. %Figure \ref{fig:r_det} shows the detection rate of Pop.~III BBHs by the Einstein Telescope (in the 10 km triangle configuration, Section~\ref{sec:r_det_method}) 
Figure \ref{fig:r_det} shows the detection rate for all our models. Overall, we expect that the Einstein Telescope will detect between $10$ and $10^4$ BBH mergers from Pop.~III stars  in one year of observation. The highest predicted rates ($\mathcal{R}_{\rm det}>10^3$  yr$^{-1}$) are associated with the SFRD from \citetalias{jaacks2019} and with the initial orbital period distribution from \citetalias{sana2012}. These numbers refer to mergers that happen inside the Einstein Telescope instrumental horizon, without distinguishing between high- and low-redshift mergers.

%We expect that a large fraction of the detected Pop.~III BBH mergers occur at redshift $z>8$, during or before the cosmic reionization. This value significantly depends on the adopted SFRD model: we estimate that 51--65\% detectable Pop.~III BBH mergers take place at $z>8$ according to \citetalias{hartwig2022}, 58--73\% according to \citetalias{skinner2020}, 41--52\% with the SFRD by \citetalias{jaacks2019}, and 23--30\% assuming \citetalias{liu_bromm_20}. These high-redshift detections are particularly important, because they will allow us to characterize the properties of Pop.~III BBHs.

We expect that a large fraction of the detected Pop. III BBH mergers occur at redshift $z > 8$, during or before the cosmic reionization. This value significantly depends on the adopted SFRD model: we estimate that 60--71\% detectable Pop.~III BBH mergers take place at $z>8$ according to \citetalias{hartwig2022}, 66-78\% according to \citetalias{skinner2020}, 45-54\% with the SFRD by \citetalias{jaacks2019}, and 28-34\% assuming \citetalias{liu_bromm_20}. These high-redshift detections are particularly important, because they will allow us to characterize the properties of Pop.~III BBHs.

%minimum =  51.46088465830039 maximum =  65.2072207944105 with  asloth_smooth_off
%minimum =  58.41344014241211 maximum =  73.22494831945578 with  SW20_smooth_off
%minimum =  40.569634624307696 maximum =  51.60399482692858 with  jaacks
%minimum =  23.087559879500223 maximum =  30.202785528328096 with  LiuBromm

\subsection{Evolution of BH mass with redshift}
\label{sec:mass_evol} 

%%%%%%%%%%%%%%%%%FIGURE%%%%%%%%%%%%%%%%%%%%%%%%
\begin{figure}
    \centering
   \includegraphics[width = \columnwidth]{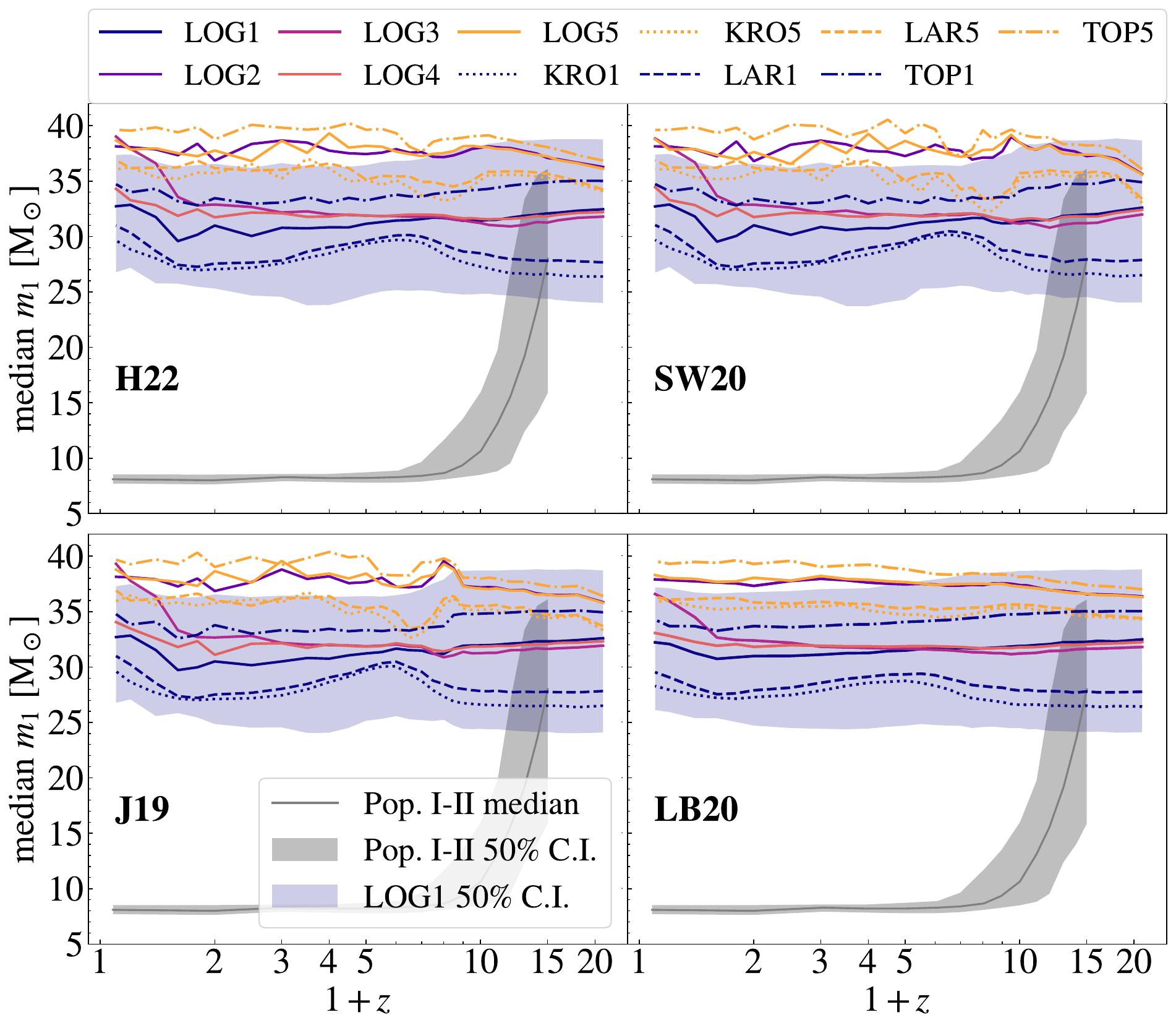}
    \caption{Median primary BH mass $m_1$ as a function of redshift, for all the models considered in this work. The blue shaded area is the interval from the 25th to the 75th percentile of the primary BH mass distribution at fixed redshift  for the LOG1 model. Upper left-hand panel: \citetalias{hartwig2022} SFRD model, upper right-hand panel: \citetalias{skinner2020}, lower left-hand panel: \citetalias{jaacks2019}, and lower right-hand panel: \citetalias{liu_bromm_20}. %The green dotted vertical line is the minimum redshift ($z_{\rm min}$) for Pop.~III star formation. 
    The grey thin solid line shows the median primary mass of Pop. I-II BBHs in our fiducial model (Appendix~\ref{sec:appendix}). The shaded grey area is the corresponding 50\% credible interval.
    } 
    \label{fig:mass0}
\end{figure}
%%%%%%%%%%%%%%%%%%%%%%%%%%%%%%%%%%%%%%%%%%%%%%

%%%%%%%%%%%%%%%%FIGURE%%%%%%%%%%%%%%%%%%%%%%%%%
\begin{figure}
    \centering
    \includegraphics[width = \columnwidth]{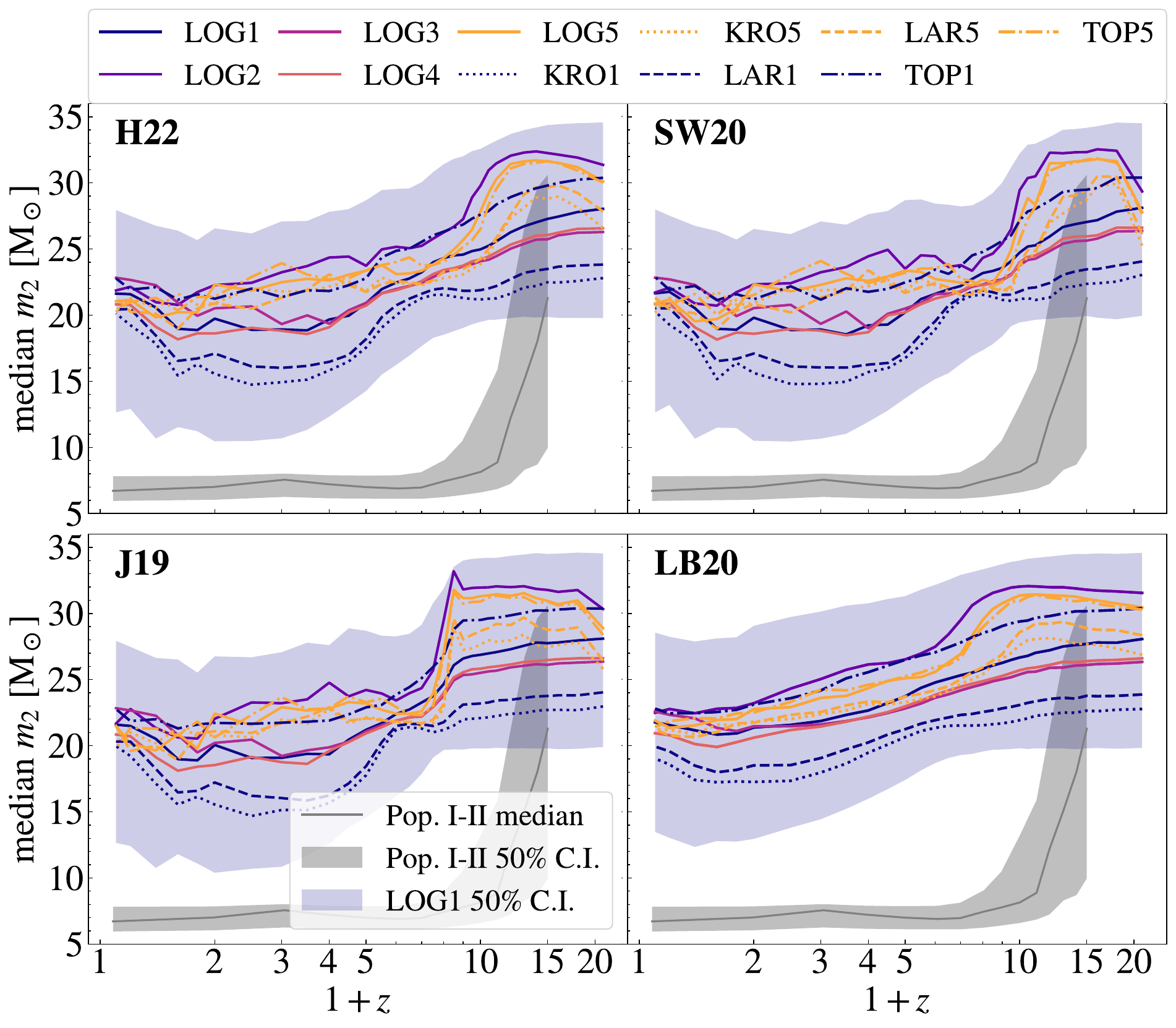}
    \caption{Same as Figure \protect \ref{fig:mass0} but for the secondary BH mass $m_2$.}
    \label{fig:mass1}
\end{figure}
%%%%%%%%%%%%%%%%%%%%%%%%%%%%%%%%%%%%%%%%%%%%%%%

%%%%%%%%%%%%FIGURE%%%%%%%%%%%%%%%%%%%%%%%%%
\begin{figure}
    \centering
    \includegraphics[width = \columnwidth]{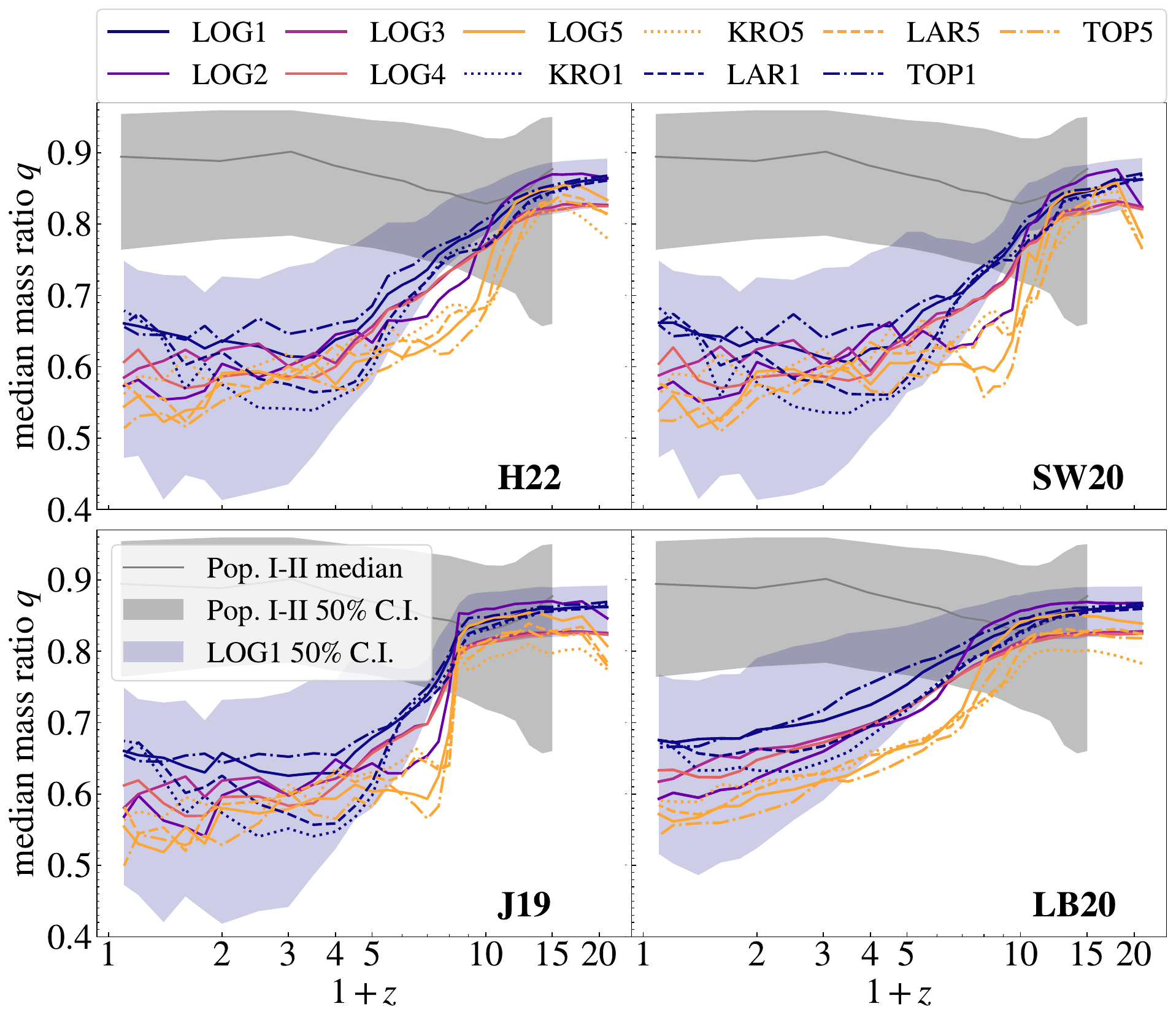}
    \caption[]{Same as Figure \protect \ref{fig:mass0} but for the BH mass ratio $q$.}
    \label{fig:mass_ratio}
\end{figure}
%%%%%%%%%%%%%%%%%%%%%%%%%%%%%%%%%%%%

%%%%%%%%%%%%%%%%%%FIGURE%%%%%%%%%%%%%%%%%%%%%%%%%
\begin{figure*}
    \centering
    \includegraphics[width = 0.8\textwidth]{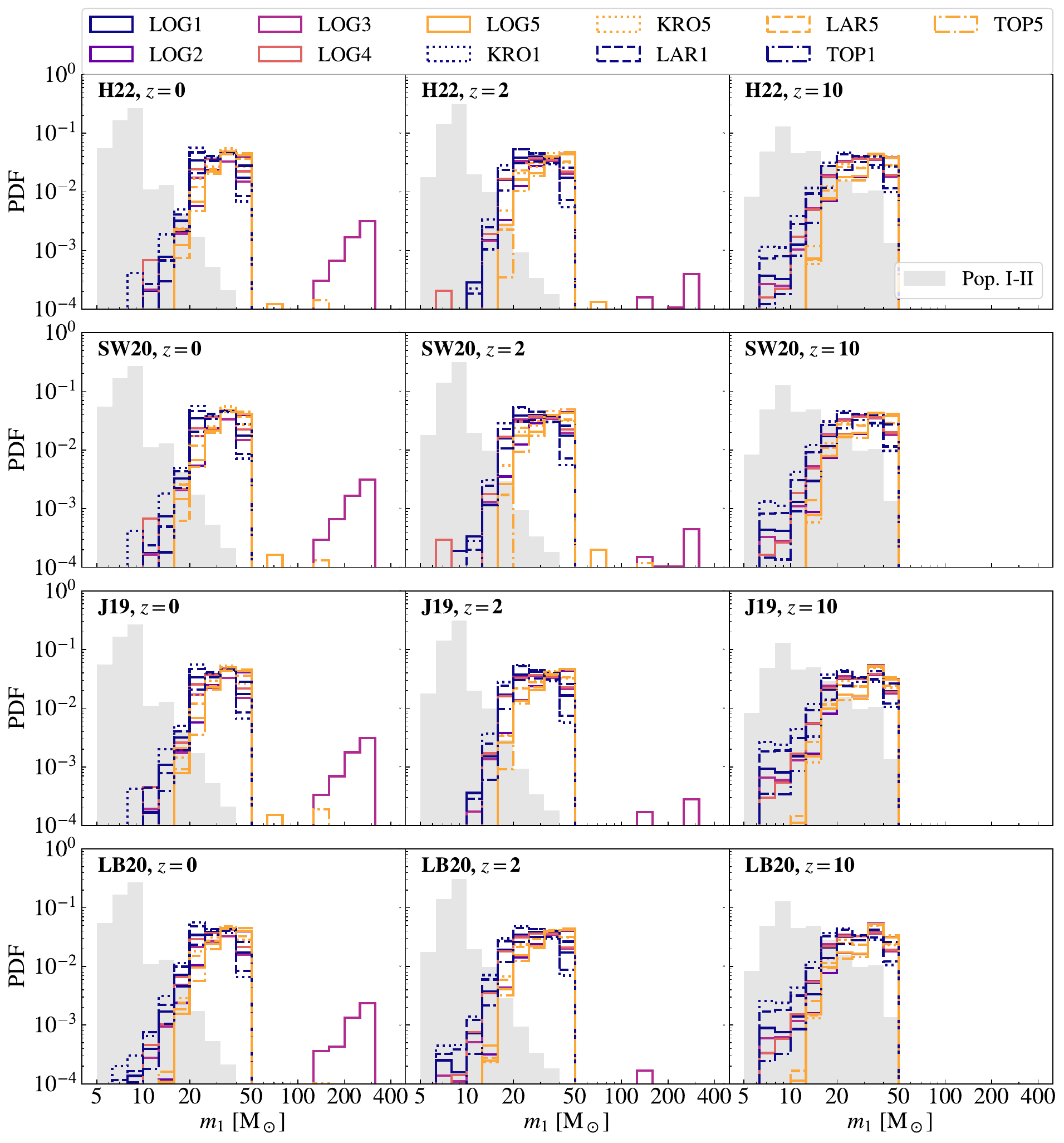}
    \caption{Primary BH mass distribution $m_1$ for three different redshift bins (from left to right $z = 0,2$ and 10) and the four SFRD models considered in this work (from top to bottom  \citetalias{hartwig2022}, \citetalias{skinner2020}, \citetalias{jaacks2019} and \citetalias{liu_bromm_20}). The grey shaded histograms show the primary mass distribution of Pop. I-II BBHs in our fiducial model (Appendix~\ref{sec:appendix}).} 
    \label{fig:mass0_dist}
\end{figure*}
%%%%%%%%%%%%%%%%%%%%%%%%%%%%%%%%%%%%%%%%%%%%%%%%%%

%%%%%%%%%%%%%%%FIGURE%%%%%%%%%%%%%%%%%%%%%%%%%%%
\begin{figure*}
    \centering
    \includegraphics[width = 0.8\textwidth]{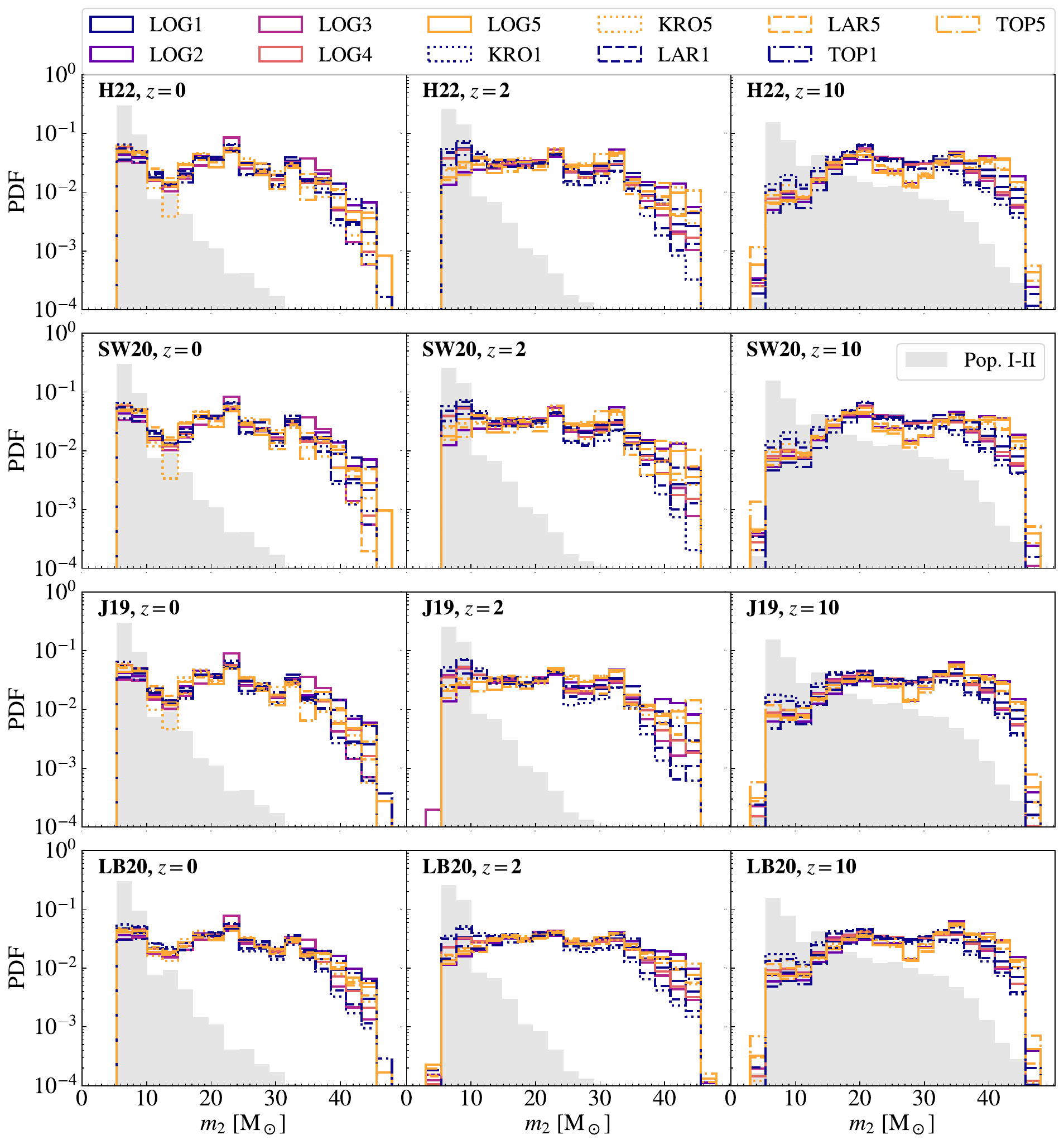}
    \caption{Same as Figure \ref{fig:mass0_dist} but for the secondary BH mass $m_2$.}
    \label{fig:mass1_dist}
\end{figure*}
%%%%%%%%%%%%%%%%%%%%%%%%%%%%%%%%%%%%%%%%%%%%%%%%%

Figure \ref{fig:mass0} shows that the median value of the primary BH mass (i.e., the most massive between the two merging BHs) does not change significantly with redshift, considering the entire ensemble of our models. On the other hand, the median mass of the secondary BH mass does decrease at lower redshift (Figure~\ref{fig:mass1}). This trend of the secondary BH mass is more evident when the SFRD of Pop.~III becomes negligible.

As a consequence, the mass ratio of  Pop.~III BBHs (Figure \ref{fig:mass_ratio}) decreases from $q \gtrsim 0.9$ at $z \sim 15$ to $q \sim 0.5 - 0.7$ at $z\leq{4}$. In contrast, the mass ratio of Pop.~I--II BBHs (Appendix \ref{sec:appendix}) remains nearly constant $q \gtrsim 0.9$ across all redshifts.

Figures \ref{fig:mass0_dist} and \ref{fig:mass1_dist} show the whole distribution of primary and secondary masses of Pop.~III BHs, respectively, at redshift $z=0,$ 2, and 10. The  percentage of Pop.~III BBHs with $m_2 \ge 25$ M$_\odot$ is $\sim{60-80}$\%  at $z=10$ and only $\sim{25-40}$\% at $z=0$ (depending on the chosen model). This change of the shape  is a result of the different distribution of delay times. 
In fact, when the formation rate of  Pop.~III stars becomes negligible, we expect to see  only mergers of Pop.~III BBHs with long delay times. We further discuss this feature in Section \ref{sec:form_channel}, considering the impact of the various formation channels.

Figures~\ref{fig:mass0}-\ref{fig:mass1_dist} also compare the behaviour of Pop.~III BBHs with that of Pop.~I-II BBHs (grey shaded area,  modelled as described in Appendix~\ref{sec:appendix}). 
Both the median primary and secondary BH mass of BBHs born from Pop.~I--II stars decrease at redshift $z<10$. 
This happens because at low redshift most BBH mergers have metal-rich progenitors, and the BH mass strongly depends on the assumed metallicity %spread 
\citep{santoliquido2021,santoliquido2022}. 

At low redshift, our intermediate- and high-metallicity stars
%metal-rich 
($Z\geq{}10^{-3}$) %binary stars 
produce a main peak in the primary BH mass distribution at $8-10$ M$_\odot$, similar to the main peak inferred from the LIGO--Virgo--KAGRA (LVK) data \citep[][]{abbottO2popandrate,abbotO3apopandrate,abbottO3bpopandrate,farah2023,callister2023}. Instead, primary BHs born from Pop.~III stars have a preference for a mass $m_1\approx{30-35}$ M$_\odot$, which is in the range of the secondary peak inferred from the LVK data \citep[][]{abbottO2popandrate,abbotO3apopandrate,abbottO3bpopandrate,farah2023,callister2023}. The secondary peak has  usually been interpreted as a signature of the pair-instability mass gap, but recently this interpretation has been put into question because the lower edge of the gap should be at higher masses \citep[$>50$ M$_\odot$, e.g.,][]{farmer2020,costa2021,woosley2021,vink2021,farag2022}.
Our results indicate that the secondary peak at $\sim{35}$ M$_\odot$ might rather be a signature of the progenitor's metallicity: metal-poor and metal-free stars in tight binary systems tend to end their life as naked helium cores with a mass of $\sim{30-40}$ M$_\odot$, favouring a sub-population of BBHs in this mass range \citep[e.g.,][]{mapelli2013,kinugawa2014,ziosi2014,belczynski2016,mapelli2016,iorio22}.

\section{Discussion}\label{sec:discussion}

\subsection{Formation channels}
\label{sec:form_channel}

%%%%%%%%%%%%%%%%FIGURE%%%%%%%%%%%%%%%%%%%%%%%%%%
\begin{figure*}
    \centering
    \includegraphics[width = 0.8\textwidth]{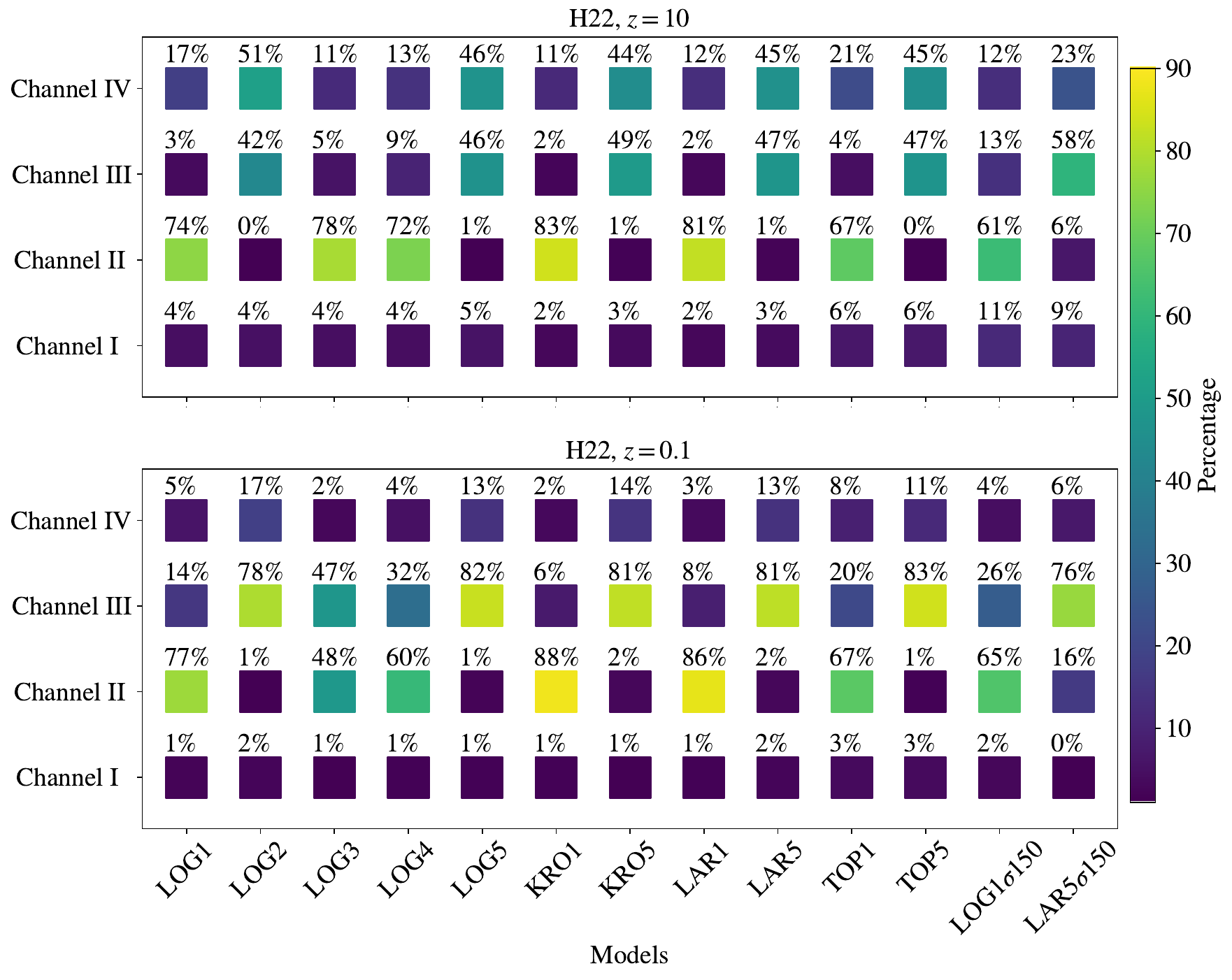}
    \caption{Percentage distribution of formation channels for all the models adopted in this work. Upper (lower) panel: Pop.~III BBHs that merge at $z=10$ ($z=0.1$).  Channel I includes all the systems that undergo a stable mass transfer before the first BH forms, and later evolve through at least one common-envelope phase. Channel II encompasses  systems that interact only via stable mass transfer (no common envelopes). Channels III and IV consist in systems that experience at least one common envelope before the formation of the first BH. The only difference between them is that one of the two stars retains a fraction of its H-rich envelope until the formation of the first BH in channel III, while both stars have lost their envelope by the formation of the first BH in  channel IV. 
    }
    \label{fig:channels}
\end{figure*}
%%%%%%%%%%%%%%%%%%%%%%%%%%%%%%%%%%%%%%%%%%%%%%%%%

To better understand the behaviour of Pop.~III BBHs, we divide our sample into the four formation channels we already discussed in \cite{iorio22}. Channel I includes all the systems that undergo a stable mass transfer before the first BH formation, and later evolve through at least one common-envelope phase. Channel II comprises the systems that interact  through at least one stable mass transfer episode, without common envelopes. 

Channel III and IV both include systems that undergo at least one common-envelope before the formation of the first BH. The only difference between these two channels is that 
 in channel III one of the two components of the binary system still retains a residual fraction of its H-rich envelope at the time of the first BH formation, while in channel IV both stars have already lost their H-rich envelope at the time of the first BH formation. %\textcolor{red}{\textbf{
 These four channels do not encompass all possible formation pathways of BBHs, but only the most common channels in our models. %}}
 
 Figure~\ref{fig:channels} shows the percentage of Pop.~III binary stars evolving through each of the four channels and resulting in BBH mergers at $z=10$ (upper panel) and $z=0.1$ (lower panel).  Channel I, which is commonly believed to be the main formation pathway for BBH mergers \citep[e.g.,][]{Tauris06,neijssel2019,belczynski2020,Mandel20,broek21}, has marginal importance ($\leq{}7$\%) for Pop.~III BBHs, regardless of the chosen initial conditions. This happens because mass transfer tends to remain stable in the late evolutionary stages, when the system is composed of a BH and a companion star, given the low mass ratio between the donor star and the BH. 

Channel II (stable mass transfer) is the dominant channel ($\geq{}50\%$) for most of our initial conditions, with the exception of LOG2, LOG5, KRO5, LAR5, and TOP5, for which channel II represents only $1-2$\% of all the mergers. The  latter five models are the only ones in our sample adopting the \citetalias{stacy2013} distribution for the orbital periods, which are significantly longer than the \citetalias{sana2012} orbital periods.  
 Indeed, the stable mass transfer channel favours high-mass binary stars that start with a short orbital separation ($\leq{}10^3$~R$_\odot$)  and undergo a stable mass transfer early in their main sequence or Hertzsprung-gap phase  \citep[e.g.,][]{Pavlovskii17,vandenheuvel17,giacobbo2018a,neijssel2019,Mandel20,marchant2021,gallos_garcia2021}. This predominance of the stable mass transfer in the case of Pop.~III stars is in agreement with \cite{kinugawa2016} and \cite{inayoshi2017}. The large orbital periods in \citetalias{stacy2013} suppress this channel, because they prevent the formation of binary systems with initial orbital separation $<10^3$~R$_\odot$.

Channel~III and IV are complementary to channel~II: they contribute together to $\sim{9-47}$\% %$\sim{8-25}$\% 
of the BBH mergers when channel~II is the dominant one and to $>90$\% when channel~II is suppressed, i.e. for models LOG2, LOG5, KRO5, LAR5, and TOP5. In the latter five cases, the initial orbital periods are sufficiently large that the two stars start mass transfer only when their radii are significantly expanded, i.e. in the red giant phase. Because of their convective envelope, such mass transfer becomes unstable and triggers one or more common envelopes. 

Channel~III has generally longer delay times than channel~IV (Figure~\ref{fig:delaytime_PopIII}) because it takes place in systems with low initial mass ratios: %[Costa et al. 2023, in prep]: \citep{costa2023}: 
the secondary star is generally less massive than the BH produced by the primary star and mass transfer episodes after the formation of the first BH do not shrink the orbit.  The long delay times of channel~III explain why it becomes more important at $z=0.1$ with respect to the high redshift.

\subsection{The evolution of the secondary mass}
\label{sec:m2_evol}

Our models calculated with the natal-kick distribution by \citetalias{giacobbo2020}  show that the median secondary mass of Pop.~III BBH mergers in the local Universe is  significantly lower than that of Pop.~III BBHs in the early Universe. This leads to a sub-population of unequal-mass BBHs ($q\sim{0.1-0.7}$), which might help us to identify Pop.~III BBHs among the other LVK mergers, since most BBHs born from Pop.~I--II stars  
are nearly equal mass in our simulations \citep{santoliquido2021,iorio22}. Also, most LVK systems are nearly
equal mass \citep{abbottO3bpopandrate}.

\begin{figure}
    \centering
    \includegraphics[width =  \columnwidth]{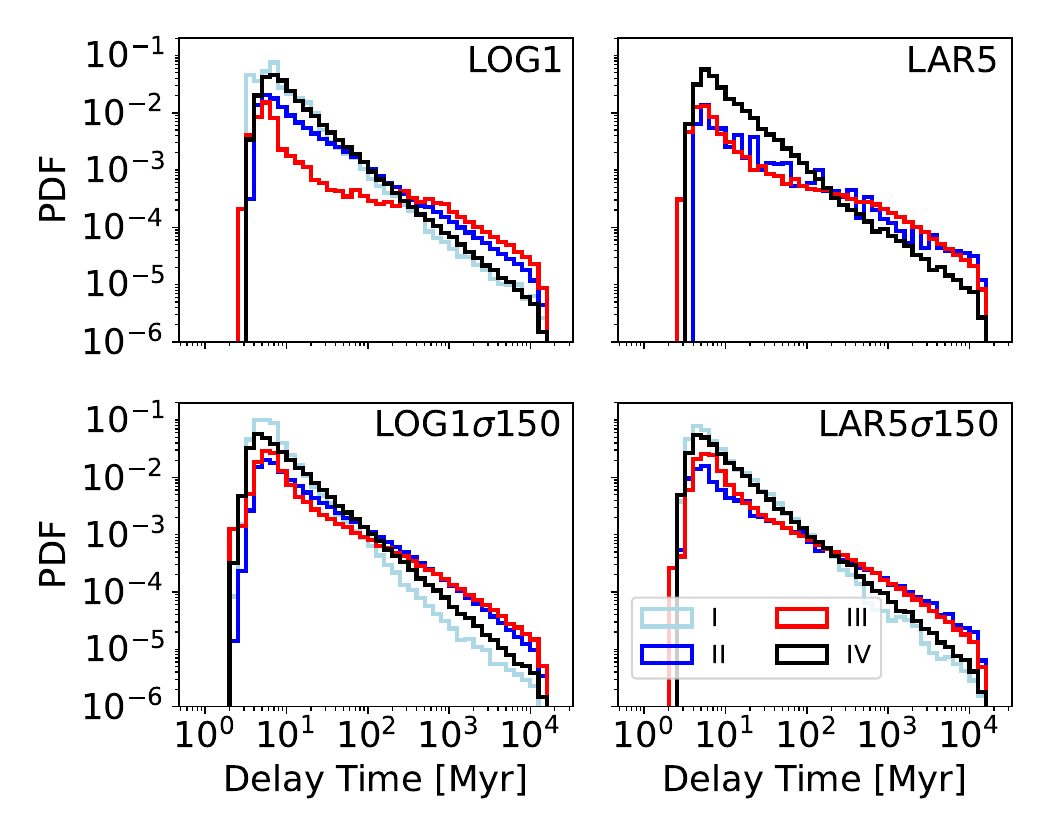}
    \caption{Distribution of delay times for models LOG1, LAR5, LOG1$\sigma{}150$ and LAR5$\sigma{}150$. Light-blue line: channel~I; blue line: channel~II; red line: channel~III; black line: channel~IV. Channel~I is not shown in the case of LAR5 because of the low number of systems. These data come directly from the \sevn{} catalogues and are not convolved with redshift evolution.}
    \label{fig:delaytime_PopIII}
\end{figure}

\begin{figure*}
    \centering
    \includegraphics[width = 0.9 \textwidth]{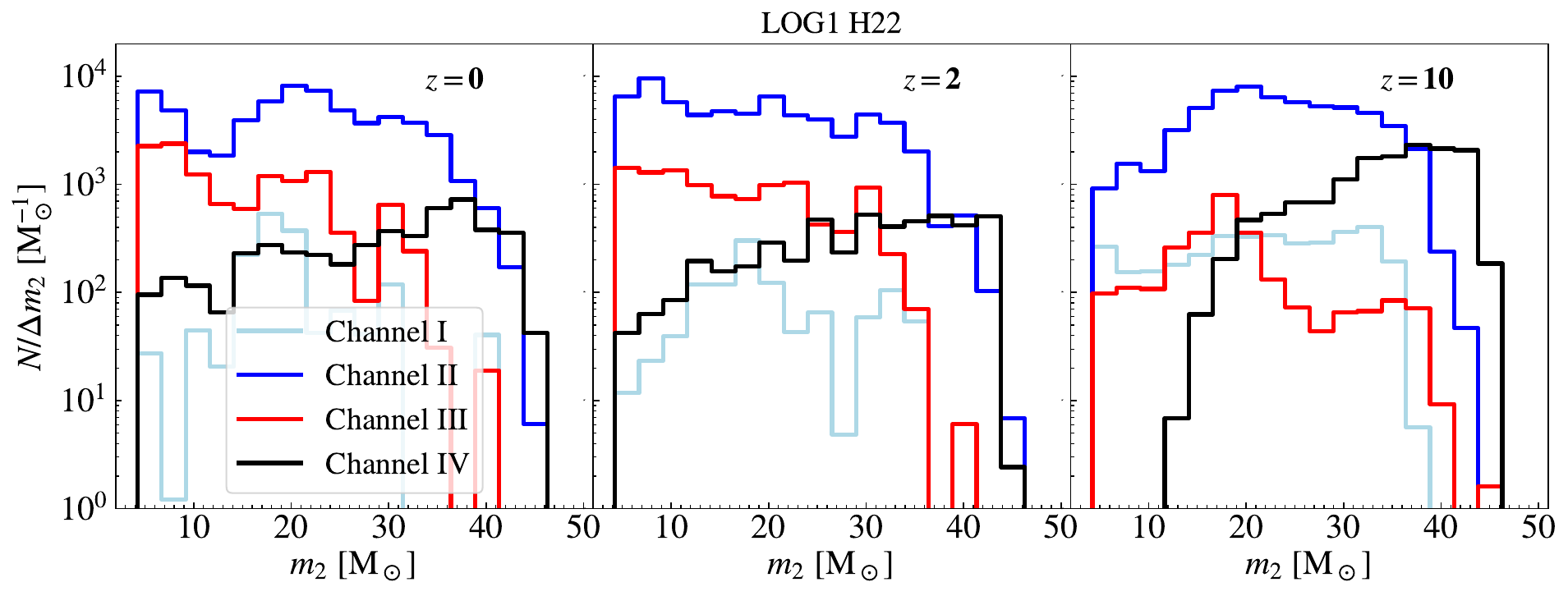}
    \caption{Secondary BH mass distribution $m_2$ for three different redshift bins (from left to right: $z = 0,\,{}2$, and 10). We show model LOG1 with the \citetalias{hartwig2022}  SFRD.  Light-blue line: channel~I; blue line: channel~II; red line: channel~III; black line: channel~IV. 
    }
    \label{fig:m2_splitchannel}
\end{figure*}

Figures \ref{fig:delaytime_PopIII} and \ref{fig:m2_splitchannel} show that this trend is an effect of delay time: the majority of the unequal mass BBHs (i.e., with low-mass secondary BHs) come from channel~II and III. These Figures show that BBHs with low-mass secondary BHs have longer delay times in  both channel~II and III.

%%%%%%%%%%%%%%FIGURE%%%%%%%%%%%%%%%%%
%%%%%%%%%%%%%%%%%%%%%%%

%%%%%%%%%%%%FIGURE%%%%%%%%%%%
%%%%%%%%%%%%%%%%%%%%%%%

\subsection{The effect of natal kicks}
\label{sec:nat_kicks}

All the models we discussed so far adopt the natal kick model from \citetalias{giacobbo2020}. This is our fiducial kick model because it naturally accounts for the claimed lower kicks in stripped and ultra-stripped supernovae \citep[e.g.,][]{bray2016,tauris2017,kruckow2018,bray2018}. On the other hand, the model by \citetalias{giacobbo2020} has a major impact on the formation channels, because it introduces a dependence of the kick on the BH and ejecta mass. 

Here, we consider an alternative model $\sigma{}150$ (Section~\ref{sec:sevn}),  in which the natal kicks have been randomly drawn from a Maxwellian distribution with parameter $\sigma{}=150$~km~s$^{-1}$. In this alternative model,  the natal kicks do not depend on the properties of the system. This implies that the $\sigma{}150$ kicks are generally  larger for stripped/ultra-stripped binaries and for high-mass BHs than the \citetalias{giacobbo2020} kicks. This difference has a substantial impact on channel~III.

Figure~\ref{fig:delaytime_PopIII} shows the delay time distribution for four models: LOG1 and LAR5 adopt the kick distribution by \citetalias{giacobbo2020}, while  LOG1$\sigma{}150$ and LAR5$\sigma{}150$  adopt the $\sigma{}150$ model.  We distinguish the delay times of the four channels. The kick model barely affects channels~II and IV, while it has a strong impact on channels~I and III. Model $\sigma{}150$ slightly increases the number of channel~I BBH mergers, from nearly 0 to a few per cent. Most importantly, model $\sigma{}150$ wildly changes the delay-time distribution of channel~III mergers, populating the region of short delay times.  

This difference springs from the impact of the natal kick on the orbital eccentricity. A larger kick either splits the binary, or increases its orbital eccentricity. Since the time of GW decay $t_{\rm GW}\propto{}(1-e^2)^{7/2}$ \citep{peters1964}, a large eccentricity speeds up the BBH merger significantly. This effect is particularly important for channel~III BBHs because they start from a large initial semi-major axis of the progenitor binary ($10^2-10^5$ R$_\odot$) and have lower secondary BH masses than the other channels \citep{costa2023}. %[Costa et al. 2023, in prep.].

\begin{figure}
    \centering
    \includegraphics[width = \columnwidth]{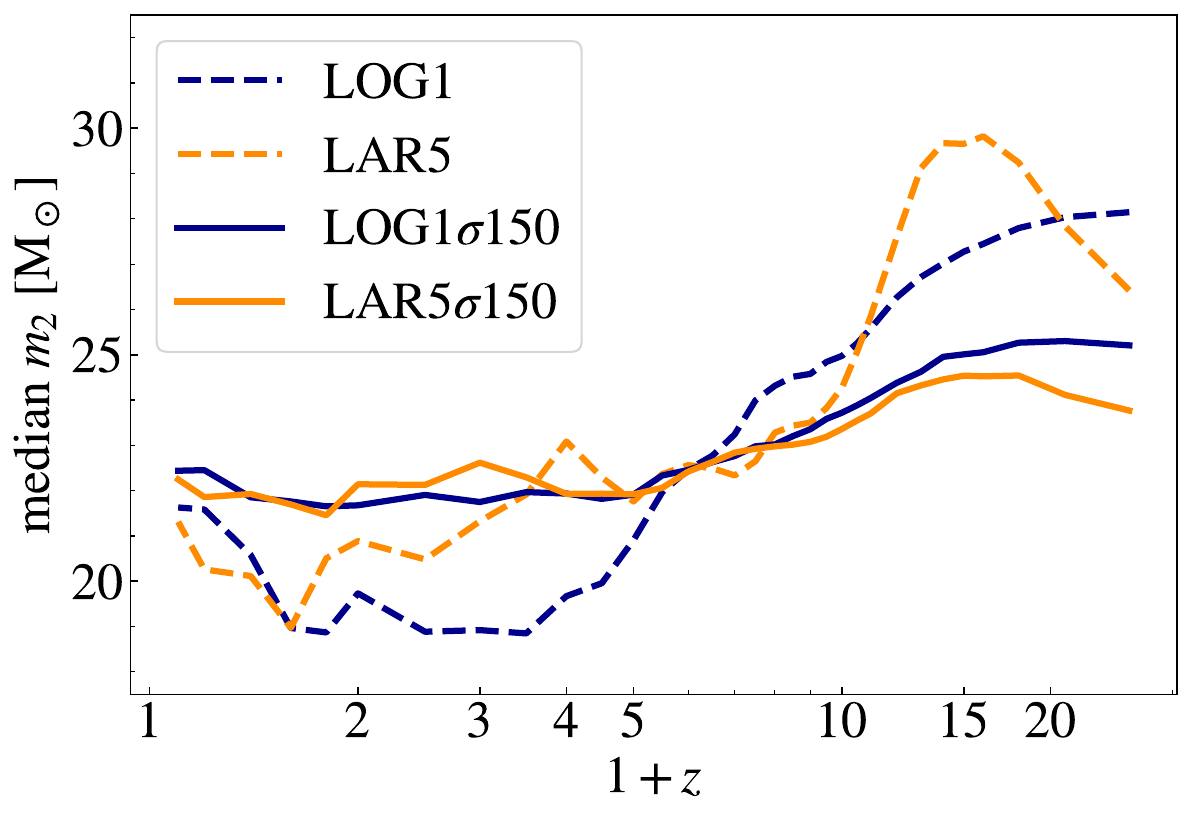}
    \caption{Evolution of the median secondary BH mass $m_2$ as a function of redshift, for LOG1 and LAR5, with the \citetalias{hartwig2022} star formation rate. Solid (dashed) line: natal kicks drawn from model $\sigma{}150$ (\citetalias{giacobbo2020}). 
    }
    \label{fig:med_m1_kick}
\end{figure}

The different delay time distribution of channel~III has an obvious impact on the median mass of the secondary BH. Figure~\ref{fig:med_m1_kick} shows the evolution of the median secondary BH mass for models LOG1, LAR5, LOG1$\sigma150$, and LAR5$\sigma{}150$ (the same models as in Figure \ref{fig:delaytime_PopIII}). 
 The decrease of the median secondary BH mass with redshift almost completely disappears in the models with $\sigma{}150$, because of the larger number of channel~III mergers and of their different delay time distribution.  
 
\begin{figure}
    \centering
    \includegraphics[width = \columnwidth]{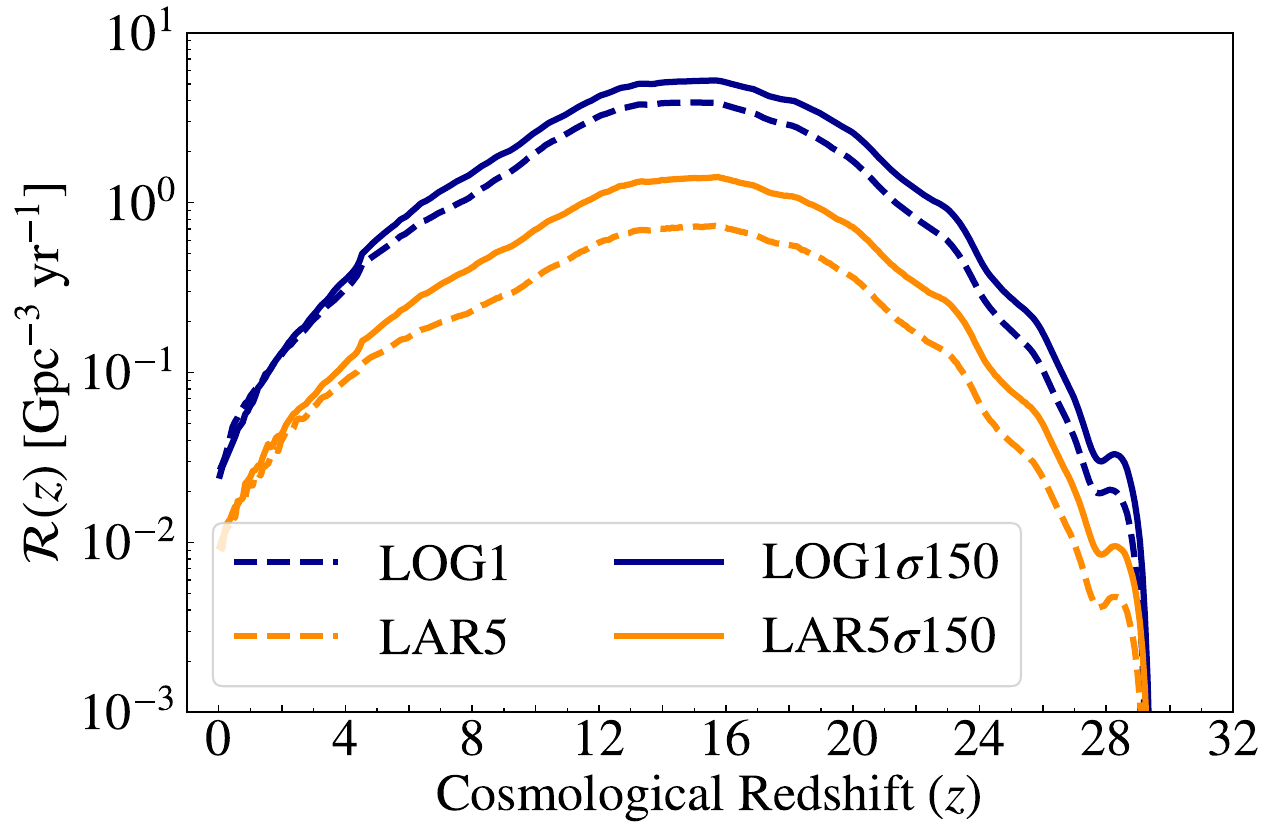}
    \caption{Evolution of the merger rate density with redshift $\mathcal{R}(z)$ for LOG1 and LAR5, with the \citetalias{hartwig2022} star formation rate. Solid (dashed) line: natal kicks from model $\sigma{}150$ (\citetalias{giacobbo2020}).}
    \label{fig:mrd_kick}
\end{figure}

 Still, the SN kick does not have a large impact on the merger rate density evolution of Pop.~III BBHs, as shown in Fig.~\ref{fig:mrd_kick}. The merger rate density of model LAR5$\sigma{}150$ is higher by a factor of two at the peak redshift ($z=13$), where the variation of the number of channel~III mergers is larger.

%%%%%%%%%%%%%%FIGURE%%%%%%%%%%%%%%%%%
%%%%%%%%%%%%%%%%%%%%%%%

%%%%%%%%%%%%%%FIGURE%%%%%%%%%%%%%%%%%

\subsection{BBH mergers above and inside the mass gap}

%\textcolor{red}{\textbf{
In our models of single Pop.~III star evolution, the pair-instability mass gap extends from a BH mass $\approx{90}$ M$_\odot$ to $\approx{240}$ M$_\odot$, because we assume that the residual hydrogen envelope of the progenitor star is not ejected during core collapse (Figure~5 and Section 3.1 of \citealt{costa2023}). %; see also  \citealt{farrell2021,costa2021,vink2021,tanikawa2021}). 
Instead, %if we 
had we assumed that the hydrogen envelope is completely ejected, the pair instability mass gap would have shifted between a BH mass $\approx{50}$ and $\approx{130}$ M$_\odot$, corresponding to the helium core mass at the boundaries of the gap. %}}

%\textcolor{red}{\textbf{
Our models of binary star evolution allow the formation of BHs with mass both above and inside the pair-instability mass gap. In binary systems, mass accretion and stellar collisions possibly populate the mass gap, because they trigger the formation of stars with undersized He cores with respect to the hydrogen-rich envelope \citep{spera2019,dicarlo2019,Renzo2020,kremer2020,tanikawa2021,banerjee2022,Costa2022,Ballone2023}. However, in our Pop.~III star simulations, BBH mergers with primary mass inside or above the gap are extremely rare. %}}

\begin{figure}
    \centering
    \includegraphics[width =  \columnwidth]{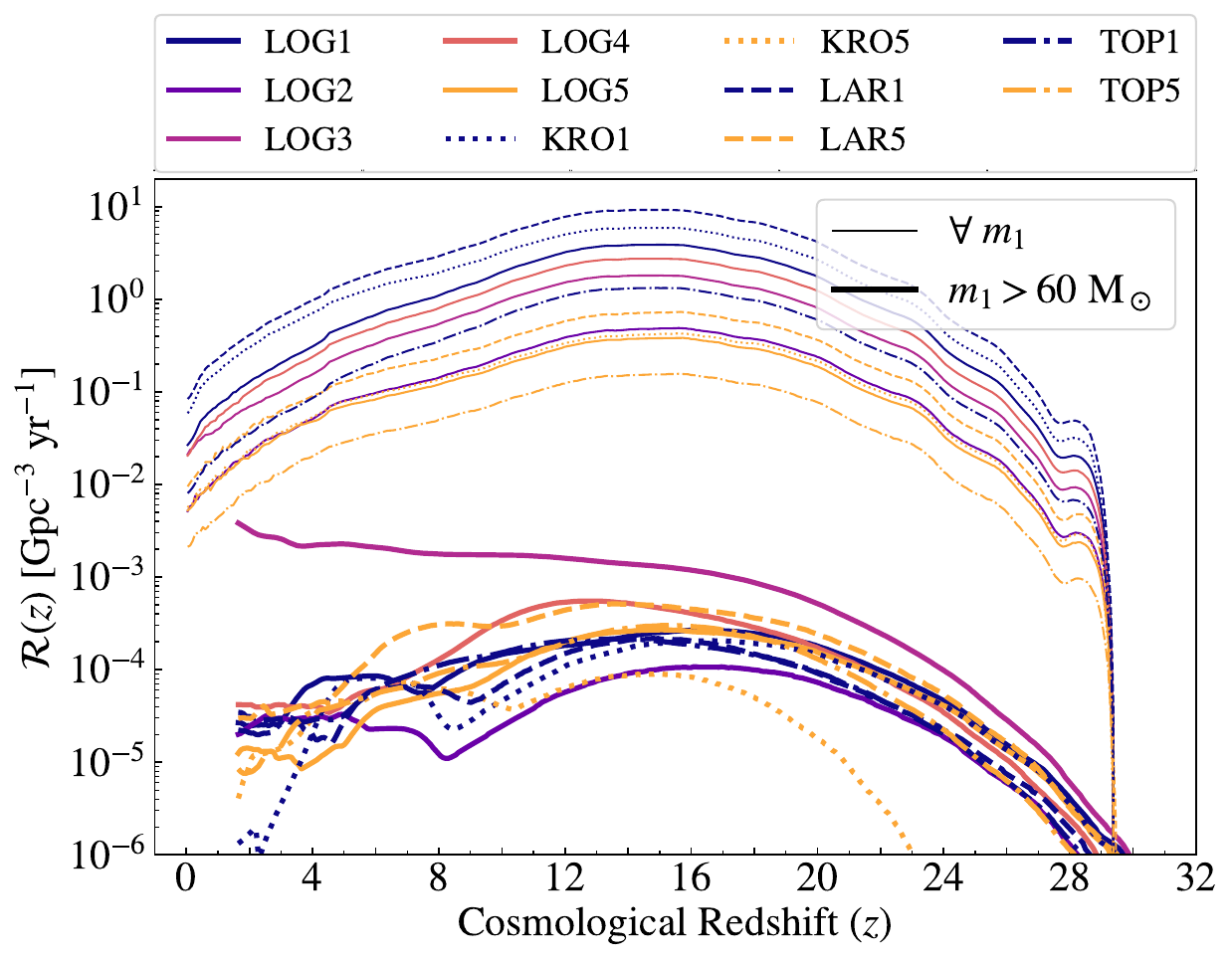}
    \caption{
    The thick lines show the merger rate density evolution of Pop.~III BBHs  with primary BH mass $m_1 > 60$ M$_\odot$. For comparison, the thin lines show the total merger rate density evolution of Pop.~III BBHs (for any value of $m_1$). For all the models in this Figure, we use the Pop.~III star SFRD from \citetalias{hartwig2022} (Figure \protect\ref{fig:sfrd}). The colours and line types refer to different initial orbital parameters (Table~\ref{tab:IC}).}
    \label{fig:popII}
\end{figure}

%\textcolor{red}{\textbf{
Figure~\ref{fig:popII} shows that our Pop.~III binary stars produce a very low merger rate density of BBHs with primary BH mass\footnote{In Figure~\ref{fig:popII}, we consider $m_1=60$ M$_\odot$ as the lower edge of the pair-instability mass gap, because this is the  most common value adopted in the literature \citep[e.g.,][]{abbottGW190521astro}, even if this value is lower than the one found in our models.} $m_1>60$~M$_\odot$ in the Local Volume [$\mathcal{R}(m_1>60\,{}\mathrm{M}_\odot,z=0)<10^{-4}$~Gpc$^{-3}$~yr$^{-1}$], apart from the LOG3 model [$\mathcal{R}(m_1>60\,{}\mathrm{M}_\odot,z=0)\approx{}4\times{}10^{-3}$~Gpc$^{-3}$~yr$^{-1}$]. In our simulations, most BHs with mass inside or above the pair-instability gap are single objects or members of loose binary systems and do not merge within the lifetime of the Universe. Dynamical interactions in dense star clusters can dramatically boost the efficiency of BBH mergers inside and above the gap, because they favour dynamical exchanges and the hardening of massive binary systems  \citep[e.g.,][]{dicarlo2020,wang2022}. %}}

%%%%%%%%%%%%%%%%FIGURE%%%%%%%%%%%%%%%%%%%%%%
%%%%%%%%%%%%%%%%%%%%%%%%%%%%%%%%%%%%%%%%%%%%z

\begin{figure}
    \centering
    \includegraphics[width = \columnwidth]{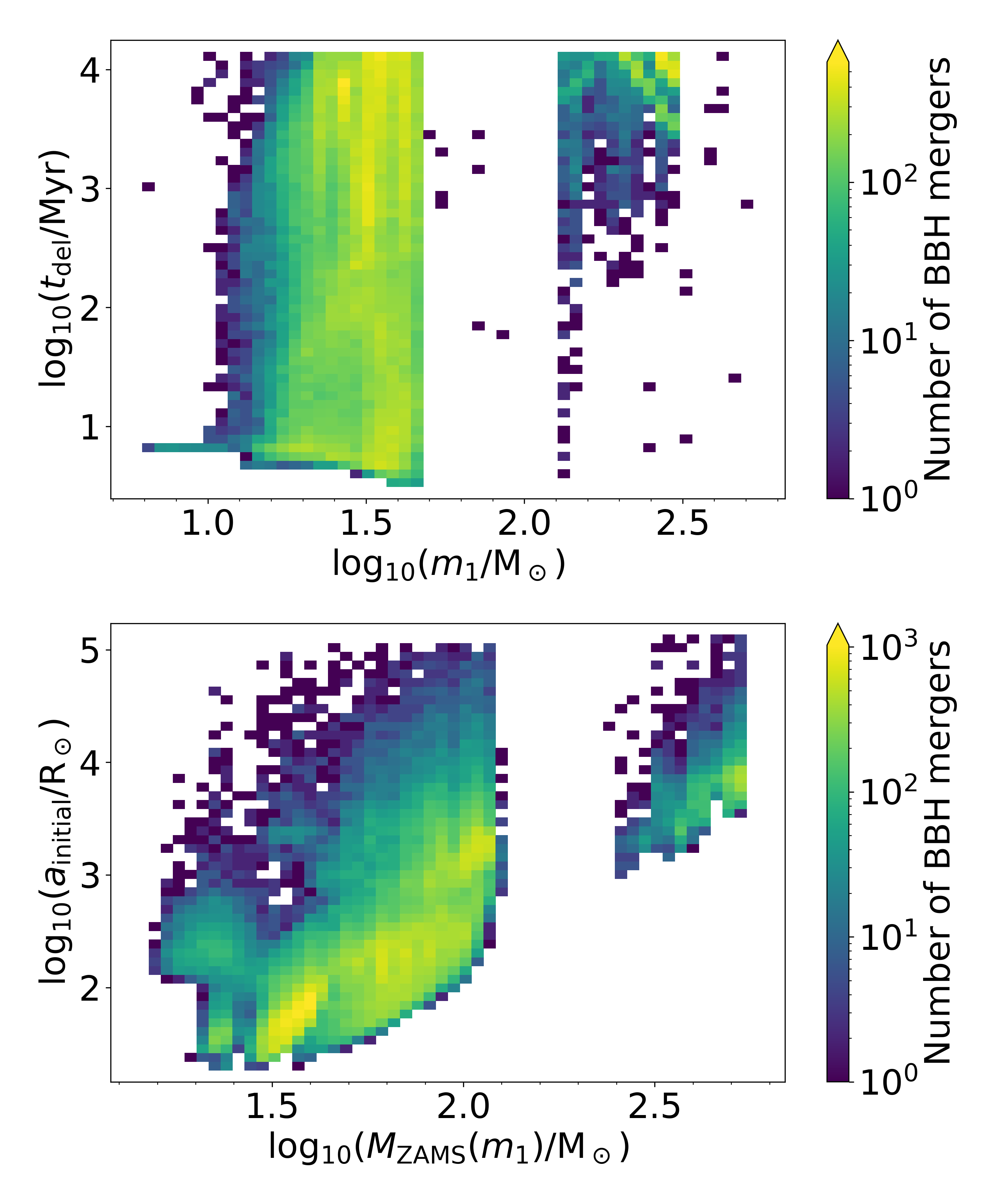}
    \caption{Properties of Pop.~III BBH mergers and their progenitors in model LOG3. Upper panel: delay time $t_{\rm del}$ as a function of the primary BH mass $m_1$. Lower panel: initial semi-major axis of the progenitor binary star $a_{\rm initial}$ versus ZAMS mass of the progenitor of the primary BH $M_{\rm ZAMS}(m_1)$. These data come directly from the \sevn{} catalogues and are not convolved with redshift evolution.
    }
    \label{fig:LOG3}
\end{figure}

%\textcolor{red}{\bf 
Furthermore, Figure~\ref{fig:mass0_dist} shows that we expect Pop.~III BBH mergers with mass above the gap only at low redshift. %}
The long delay times of channel~III explain why we have BBH mergers with primary BH mass above the mass gap ($>120$ M$_\odot$) only at low redshift in Figure~\ref{fig:mass0_dist}, mainly in model LOG3. This model is the only one adopting a sorted distribution to pair up the progenitor stars. Hence, it is the one with the lowest initial mass ratios ($M_{\rm ZAMS,2}/M_{\rm ZAMS,1}$). Systems with $M_{\rm ZAMS,1}\geq{}250$~M$_\odot$ and initial semi-major axis $a_{\rm initial}\in[10^3,1.5\times{}10^5]$ R$_\odot$ evolve nearly unperturbed until the primary star becomes a giant star and fills its Roche lobe (Figure~\ref{fig:LOG3}). In channel~III, the Roche lobe overflow becomes unstable and triggers a common envelope which removes the H-rich envelope of the primary star. Shortly  after the common envelope phase, the primary star collapses to a BH above the pair-instability mass gap. Then, the binary evolves nearly unperturbed until the secondary star also becomes a BH. Given the large semi-major axis at the time of formation of the secondary BH ($\approx{100}$ R$_\odot$) and the relatively low mass of the secondary BH ($m_2\sim{10-30}$ M$_\odot$), such binaries have a long delay time, of the order of  $5-12$ Gyr (Figure~\ref{fig:LOG3}).

%%%%%%%%%%%%%%FIGURE%%%%%%%%%%%%%%%%%
%%%%%%%%%%%%%%%%%%%%%%%

\subsection{Comparison with previous work}

We do not find any BBH mergers with primary mass in the $\sim{100-200}$ M$_\odot$ regime,  whereas \cite{tanikawa2022b} find this sub-population of mergers in their fiducial model.  This is mainly an effect of the different stellar radii. \cite{tanikawa2022b} produce this sub-population of BBH mergers from binary stars with primary ZAMS mass $\sim{65-90}$ M$_\odot$. In their fiducial model, such stars have radii $R<100$~R$_\odot$ for their entire life, while in our models they expand much more during the end of the main sequence and the red giant phase \citep{costa2023}. This  comes  from the choice of core overshooting: we assume an overshooting parameter $\Lambda_{\rm ov}=0.5$ in units of pressure scale height, which corresponds to  $f_{\rm ov}=0.025$ in the formalism adopted by \cite{tanikawa2022b}, while  they assume $f_{\rm ov}=0.01$ in the fiducial model (M-model). Indeed, our models are similar to the L-std model  by \cite{tanikawa2022b}, with $f_{\rm ov}=0.03$, and in this model they find no mergers with primary BH mass inside the mass gap (Fig.~6 of \citealt{tanikawa2022b}).

The  merger rate density of Pop.~III BBHs estimated by \cite{tanikawa2022b}  reaches a maximum of $R(z\sim{10})\approx{20}$ Gpc$^{-3}$ yr$^{-1}$. In terms of the initial orbital parameters, their model is almost identical to our LOG1 model. They use the SFRD from \citetalias{skinner2020}. In our LOG1 model with the \citetalias{skinner2020} SFRD, the merger rate peaks at $z\sim{16}$  and $R(z\sim{16})\approx{2}$ Gpc$^{-3}$ yr$^{-1}$. The difference in the redshift of the peak %{\filippo{{\st{, which is $z\sim{16}$ and 10 between us and    \protect\cite{tanikawa2022b}}}{\st{ with the \protect\citetalias{skinner2020}}}{\st{ SFRD,}}}} 
is a consequence of the delay time distribution. \cite{tanikawa2022b} have significantly longer delay times, even for their L-std model \citep[see Figure~3 from][]{tanikawa2021b}. The large difference in the normalisation of the peak between our work and \cite{tanikawa2022b} is a consequence of the differences in our stellar and binary evolution models. In particular, our larger stellar radii increase the %{\filippo{{\st{risk}} possibility
chance that two possible progenitor stars collide leaving just one single star, before they become a BBH. 

%\textcolor{red}{\bf Here, we use models of non-spinning Pop.~III stars. }
%MM: ADD discussion on quasi homogeneous evolution and different stellar evolution models TBD}

\subsection{A proxy for chemically homogeneous evolution (CHE)}\label{app:mass_CHE}

%%%%%%%%%%%%%FIGURE%%%%%%%%%%%%%%%%%
\begin{figure*}
    \centering
    \includegraphics[width = \textwidth]{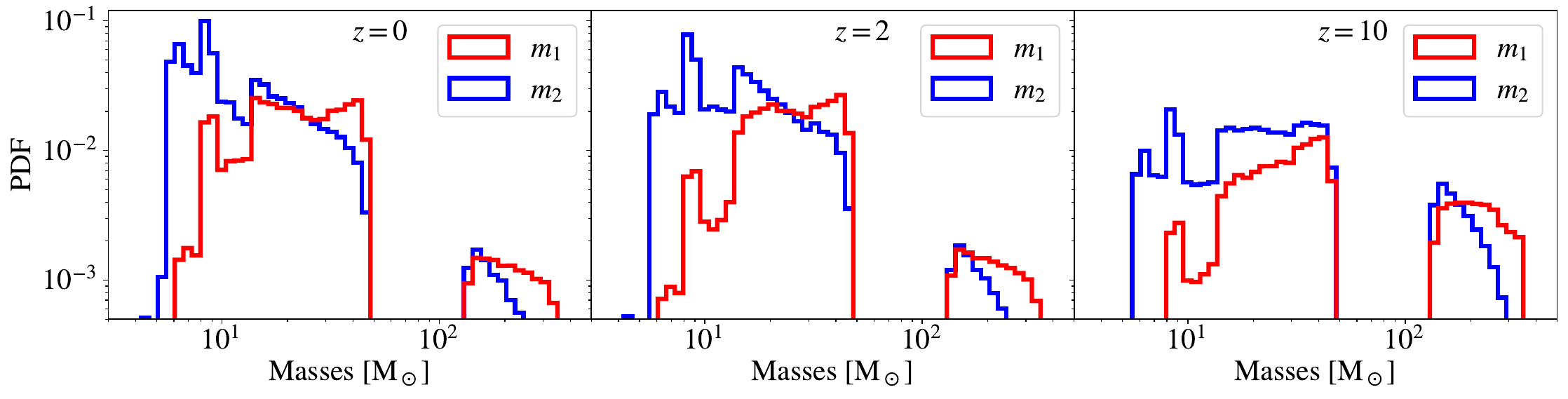}
    \caption{%\textcolor{red}{\bf 
    Primary BH mass (red) and secondary BH mass (blue) of Pop.~III BBHs merging at redshift $z=0,$ (left), 2 (middle), and 10 (right) assuming the star formation history from \protect\citetalias{hartwig2022} and the initial binary orbital parameters as in model LOG1 (Table~\ref{tab:IC}). Here, we evolve pure-helium stars. %}
    }
    \label{fig:mass_CHE}
\end{figure*}
%%%%%%%%%%%%%%%%%%%%%%%

%\textcolor{red}{\bf %In the main text, we have shown the evolution of non-spinning stars \citep{costa2023}. 
We have adopted models of non-spinning Pop.~III stars \citep{costa2023}.  
Pop.~III stars might be fast spinning \citep{yoon2012,choplin2019}. %, leading to chemically homogeneous evolution (CHE) because of low wind mass-loss \citep{demink2016,mandel2016,andrews2020,dubuisson2020}. %Our stellar-evolution models \citep{costa2023} are non spinning. 
According to \cite{tanikawa2021}, we can have a glimpse at what happens to fast-spinning stars by considering the evolution of pure-helium stars. In fact, fast spinning stars at low-metallicity effectively evolve toward chemically homogeneous evolution (CHE, \citealt{demink2016,mandel2016,marchant2016,dubuisson2020,riley2021}). %Here, 
In this discussion, we  use pure-He stars  as a simplified proxy for CHE, which we will model carefully in future work. %}

\begin{figure}
    \centering
    \includegraphics[width = \columnwidth]{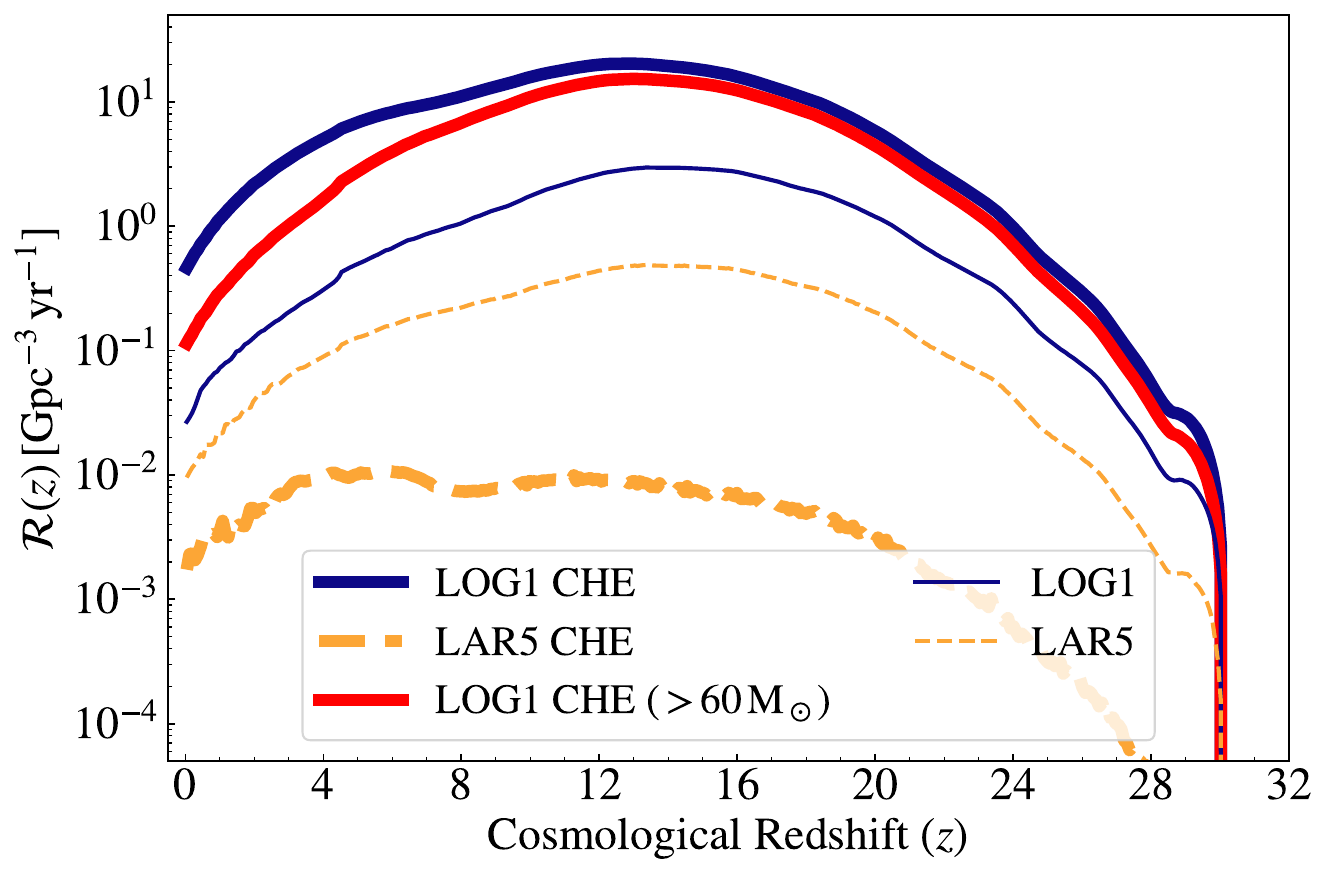}
    \caption{%\textcolor{red}{\bf 
    Merger rate density of Pop.~III BBHs born from pure-helium binary stars assuming the star formation history from \protect\citetalias{hartwig2022}. Thick solid blue line: pure-helium binary stars evolved with LOG1 initial conditions. Thick solid red line: we show only the merger rate density of BBHs with primary mass $>60$ M$_\odot$ for pure-helium binary stars evolved with LOG1 initial conditions. Thick dashed orange line: pure-helium binary stars evolved with LAR5 initial conditions. Thin solid blue (Thin dashed orange)  line: model LOG1 (LAR5) with pure-hydrogen binary stars for comparison. %}
    }
    \label{fig:rate_CHE}
\end{figure}

%\textcolor{red}{\bf 
Figure~\ref{fig:mass_CHE} shows that BBHs born from Pop.~III pure-helium stars are crucially different from the other models presented here. Since pure-helium stars evolve with small radii ($<100$ R$_\odot$), the most massive progenitor stars do not merge in the early stages of their life and efficiently produce BBH mergers above the mass gap. In the case of model LOG1, the merger rate density of Pop.~III BBHs born from pure-He stars (Fig.~\ref{fig:rate_CHE}) is higher than that of Pop.~III BBHs born from H-rich stars. Moreover, the merger rate density of BBHs with primary mass $m_1>60$~M$_\odot$ is at least three orders of magnitude higher for pure-He star progenitors.  %}

%%%%%%%%%%%%%FIGURE%%%%%%%%%%%%%%%%%
%%%%%%%%%%%%%%%%%%%%%%%
%%\textcolor{red}{\bf MM: ADD short comparison with dynamics for BHs above and inside gap}

%\textcolor{red}{\bf 
If we simulate pure-helium progenitors, we obtain similar results, i.e. a higher merger rate density and a larger population of BBHs above the gap, for all the modes assuming the \citetalias{sana2012} initial orbital periods (i.e. LOG1, LOG3, LOG4, KRO1, LAR1, and TOP1). %} \textcolor{red}{\bf 
In contrast, the merger rate density of models LOG2, LOG5, KRO5, LAR5, and TOP5 drops by at least two orders of magnitude in the case of pure-helium progenitors. The reason is that these models adopt the orbital period distribution from \citetalias{stacy2013}, which suppresses short orbital separations. Since pure-helium stars are compact throughout their entire life, most of them do not undergo Roche-lobe overflow during their life and their orbital separations remain too large to merge within the lifetime of the Universe. %}

\subsection{Pop.~III BHs versus primordial BHs}

The merger rate density of primordial BHs is predicted to scale as \citep[e.g.,][]{deluca2020,Mukherjee21,Mukherjee2021b,Mukherjee22,deluca22,ng2022,franciolini2022}
\begin{equation}
\mathcal{R}_{\rm PBH}(z)=\mathcal{R}_{\rm PBH}(0)\,{}\left(\frac{t_{\rm age}(z)}{t_{\rm age}(0)}\right)^{-34/37},
\label{eq:franciolini}
\end{equation}
where $t_{\rm age}(z)$ is the age of the Universe at redshift $z$. According to \cite{franciolini2022}, current LVK data suggest that $\mathcal{R}_{\rm PBH}(0)\leq{}5.3$ Gpc$^{-3}$ yr$^{-1}$ at 90\% credible interval, if we assume primordial BH masses $\geq{}3$ M$_\odot$. This implies that our predicted merger rate density of Pop.~III BBHs is always below the upper bound of the merger rate density of primordial BHs, even for the \citetalias{jaacks2019} model at $z\approx{8}$, for which we have $\mathcal{R}(z=8)\leq{}80$ Gpc$^{-3}$ yr$^{-1}$ (Fig.~\ref{fig:mrd_pro}) versus $\mathcal{R}_{\rm PBH}(z=8)\leq{90}$  Gpc$^{-3}$ yr$^{-1}$. 

If we assume that primordial BHs represent a fraction $10^{-4}$ of the total dark matter energy density (e.g., Fig.~3 from \citealt{ng2022}), then we expect that Pop.~III BBH mergers outnumber primordial BH mergers out to redshift $z\sim{15-20}$ in our most optimistic models. Overall, the slope of the merger rate density of primordial BHs is well constrained by models, while the uncertainties about the normalisation are  larger than the ones about Pop.~III BBHs \citep[e.g.,][]{raidal2019,vaskonen2020,deluca22}, preventing us from drawing further conclusions. As discussed in \cite{ng2022}, the Einstein Telescope and  Cosmic Explorer might be able to disentangle the two populations of mergers by looking at the overall shape of the merger rate density evolution. Additional information will come from the mass distribution of BBH mergers at high redshift \citep[e.g.,][]{franciolini2022}.

\section{Summary and conclusions}
\label{sec:summary}

We estimated the merger rate density evolution of binary black holes (BBHs) born from Pop. III stars (Figure \ref{fig:mrd_pro}) by means of our code \cosmorate{} \citep{santoliquido2021}. 
To evaluate the main uncertainties affecting the merger rate density, we explored a large portion of the parameter space, %{\filippo{{\st{consisting}} 
making use of four different models for the formation history of Pop. III stars (from  \citealt[][\citetalias{jaacks2019}]{jaacks2019}, \citealt[][\citetalias{liu_bromm_20}]{liu_bromm_20}, \citealt[\citetalias{skinner2020}]{skinner2020}, and \citealt[][\citetalias{hartwig2022}]{hartwig2022}), and eleven different configurations of the initial orbital properties of Pop.~III binary stars. In particular, we probe different IMFs (flat-in-log, \citetalias{kroupa2001}, \citetalias{larson1998}, and top-heavy), mass ratios (\citetalias{sana2012}, sorted), orbital period distributions (\citetalias{sana2012}, \citetalias{stacy2013}), and eccentricity distributions (\citetalias{sana2012}, thermal), as described in Table~\ref{tab:IC}. We generated the catalogues of Pop. III BBHs with our binary population-synthesis simulation code \sevn \citep{iorio22}, based on a new set of Pop.~III stellar tracks with metallicity $Z=10^{-11}$ and ZAMS mass $m_{\rm ZAMS}\in{}[2,\,{}600]$ M$_\odot$ \citep{costa2023}. 

The assumed star formation rate history of Pop.~III stars affects both the normalisation and the shape of the BBH merger rate density evolution with redshift (Fig.~\ref{fig:mrd_pro}): $\mathcal{R}(z)$ peaks at $z_{\rm p}\approx{8-10}$ for the models by \citetalias{jaacks2019} and \citetalias{liu_bromm_20}, and at $z_{\rm p}\approx{12-16}$ for \citetalias{hartwig2022} and \citetalias{skinner2020}. For our fiducial model LOG1, the maximum merger rate density ranges from $\mathcal{R}(z_{\rm p})\approx{30}$ Gpc$^{-3}$ yr$^{-1}$ for the star formation rate density (SFRD) by \citetalias{jaacks2019}  down to $\mathcal{R}(z_{\rm p})\approx{2-4}$ Gpc$^{-3}$ yr$^{-1}$ for the SFRDs by \citetalias{skinner2020}, \citetalias{hartwig2022}, and \citetalias{liu_bromm_20}. 

At redshift $z=0$, all the considered SFRD models yield $\mathcal{R}(0)\leq{}2\times{}10^{-1}$ Gpc$^{-3}$ yr$^{-1}$ in our fiducial model LOG1, about two orders of magnitude lower than the local BBH merger rate density inferred from LVK data \citep{abbottO3bpopandrate}. Overall, changing the SFRD model for Pop.~III stars affects the BBH merger rate density by up to about one order of magnitude. In the case of the SFRD derived by \citetalias{hartwig2022}, we can also account for the intrinsic uncertainties of the SFRD calibration on data. We find that the merger rate density changes by about one order of magnitude within the 95\% credible interval of the Pop.~III SFRD estimated by \citetalias{hartwig2022} (Fig.~\ref{fig:mrd_asloth_uncrt}). 

The initial orbital properties of our Pop.~III binary systems have an even larger impact on the BBH merger rate density, up to two orders of magnitude (Fig.~\ref{fig:mrd_pro}). The models adopting a \citetalias{stacy2013} distribution for the initial orbital periods (LOG2, LOG5, KRO5, LAR5, and TOP5) have lower merger
rate densities than models adopting the distribution by \citetalias{sana2012} (LOG1, LOG3, LOG4, KRO1, LAR1, and TOP1). The reason is that short orbital periods, as in the case of \citetalias{sana2012}, favour the merger of BBHs via stable mass transfer episodes between the progenitor stars, while large orbital periods (\citetalias{stacy2013}) suppress these systems.

We estimated the mass distribution of Pop.~III BBHs for all of our models as a function of redshift. Both the primary and secondary BH (i.e., the most and least massive member of a BBH) born from a Pop.~III binary star tend to be substantially more massive than the primary and secondary BH born from a metal-rich binary star (Figs.~\ref{fig:mass0_dist},~\ref{fig:mass1_dist}). This happens mainly because stellar winds are suppressed at low $Z$. The median mass of the primary BHs born from Pop.~III stars is $m_1\approx{30-40}$~M$_\odot$ across the entire redshift range, while the median mass of primary BHs born from metal-rich stars is $m_1\approx{8}$~M$_\odot$ (Fig.~\ref{fig:mass0}). This result does not depend on the adopted SFRD and is only mildly sensitive to the initial orbital properties of Pop.~III binary stars.

The mass spectrum of primary BHs inferred by the LVK \citep{abbottO3bpopandrate} is characterised by two  peaks, the main one at 8 -- 10 M$_\odot$ and the other at $\sim{35}$ M$_\odot$. The location of these two peaks is remarkably similar to the median mass of the primary BHs born from metal-rich and metal-free/metal-poor stars in our simulations. 

The mass ratio $q$ between the secondary and primary BH is another feature that distinguishes BBHs  born from Pop.~III and metal-rich binary stars (Fig.~\ref{fig:mass_ratio}). Pop.~III BBHs merging at low redshift ($z\leq{}4$) have low mass ratios (median values $q\approx{0.5-0.7}$) with respect to BBH mergers from metal-rich stars (median values $q\approx{0.9}$). In contrast, at high redshift even BBH mergers born from Pop.~III stars have a typical $q\sim{0.9}$. This happens because the median secondary BH mass of Pop.~III BBH mergers decreases with redshift. This feature is a consequence of the delay time distribution: Pop.~III BBHs with relatively small secondary BH mass are associated with longer delay times than Pop.~III BBHs with equal mass BHs (Fig.~\ref{fig:delaytime_PopIII}). This dependence of the delay time on the secondary BH mass is a consequence of the formation channels of our Pop.~III BBHs. It is not affected by the adopted SFRD and is only mildly sensitive to the initial orbital properties of Pop.~III binary stars, but it is highly sensitive to the assumed natal kick distribution.  In our fiducial models, we assume that natal kicks are lower for more massive BHs and for (ultra-)stripped binary systems. If we instead use a natal kick model in which the kick magnitude does not depend on the properties of the system, the decrease of the median secondary BH mass with redshift almost disappears.
 
In our fiducial model (LOG1) and  all the other models assuming the initial orbital period distribution by \citetalias{sana2012}, most ($>50\%$) of our Pop.~III BBHs evolve via stable mass transfer episodes, without common envelope phases. This happens because the mass of Pop.~III BHs is always sufficiently large with respect to the mass of the companion star to avoid common envelope \citep{costa2023}.

Even if most of our BBH mergers from Pop.~III stars are rather massive ($m_1\approx{30-40}$ M$_\odot$), BBHs with mass above or inside the pair-instability mass gap are extremely rare in our models. For example, assuming the SFRD by \citetalias{hartwig2022}, we find that the local merger rate of Pop.~III BBHs with primary BH mass $m_1>60$ M$_\odot$ is $<10^{-4}$ Gpc$^{-3}$ yr$^{-1}$ in all of our models but LOG3, for which we find $\approx{4\times{}10^{-3}}$ Gpc$^{-3}$ yr$^{-1}$. 

Altogether, we expect that the Einstein Telescope will detect between 10 and $10^4$  BBH mergers from Pop.~III stars per year, depending on the adopted parameters. %In particular, for our fiducial model (LOG1 with SFRD from \citetalias{hartwig2022}) we expect $\approx{500}$ detections per year, of which 62\% from BBH mergers occurring at redshift $z>8$. 
In particular, for our fiducial model (LOG1 with SFRD from \citetalias{hartwig2022}) we expect $\approx{530}$ detections per year, of which 68\% from BBH mergers occurring at redshift $z>8$. Since the properties of low-redshift Pop.~III BBH mergers are not dramatically different from those of BBHs originating from metal-rich stars, such high-redshift detections will be crucial to characterize the population of Pop.~III BBHs. In a follow-up study, we will run parameter estimation on our simulated systems, to verify how well we can reconstruct their properties (mass and merger redshift) with the Einstein Telescope.

Our results show that the overall uncertainty %{\filippo{{\st{on}} of}} 
on the merger rate density evolution of Pop.~III BBHs mergers spans at least two orders of magnitude and depends on the SFRD model, initial orbital properties of Pop.~III binary stars, and stellar/binary evolution physics. Future work should further explore the impact of stellar evolution (e.g., rotation, chemically homogeneous evolution, core overshooting) and different assumptions for mass and angular momentum evolution during mass transfer.

\section*{Acknowledgements}

%\textcolor{red}{\bf 
We thank the anonymous referee for their insightful comments which helped us improve this work. %}
We are grateful to  Stanislav Babak, Irina Dvorkin, Gabriele Franciolini, Cecilio García-Quirós, Tomoya Kinugawa,  Natalia Korsakova,   Paolo Pani, Federico Pozzoli, and  Ataru Tanikawa %,   \textcolor{red}{OTHERS?} 
for their enlightening comments. 
GC, GI, MM, and FS acknowledge financial support from the European Research Council (ERC) for the ERC Consolidator grant DEMOBLACK, under contract no. 770017. TH acknowledges financial support from JSPS (KAKENHI Grant Numbers 19K23437 and 20K14464) and the German Environment Agency. FS thanks the  Astroparticule et Cosmologie Laboratoire (APC) and  the Institute d'Astrophysique de Paris (IAP) for hospitality during the preparation of this manuscript.  
%
%\textcolor{red}{\bf 
RSK and SCOG acknowledge financial support from the ERC via the ERC Synergy Grant ``ECOGAL'' (project ID 855130),  from the German Excellence Strategy via the Heidelberg Cluster of Excellence (EXC 2181 - 390900948) ``STRUCTURES'', and from the German Ministry for Economic Affairs and Climate Action in project ``MAINN'' (funding ID 50OO2206). RSK and SCOG  also thank for computing resources provided by the Ministry of Science, Research and the Arts (MWK) of the State of Baden-W\"{u}rttemberg through bwHPC and the German Science Foundation (DFG) through grant INST 35/1134-1 FUGG and for data storage at SDS@hd through grant INST 35/1314-1 FUGG. %}

%%%%%%%%%%%%%%%%%%%%%%%%%%%%%%%%%%%%%%%%%%%%%%%%%%
\section*{Data Availability}
%The inclusion of a Data Availability Statement is a requirement for articles published in MNRAS. Data Availability Statements provide a standardised format for readers to understand the availability of data underlying the research results described in the article. The statement may refer to original data generated in the course of the study or to third-party data analysed in the article. The statement should describe and provide means of access, where possible, by linking to the data or providing the required accession numbers for the relevant databases or DOIs.
%{\filippo{{\st{The data underlying this article will be shared on reasonable request to the corresponding authors}} All 
The main data presented in this work are publicly available on Zenodo at \cite{santoliquido2023_zenodo_v2}. The latest public version of {\sc sevn} can be downloaded from \href{https://gitlab.com/sevncodes/sevn.git}{this repository}\footnote{\href{https://gitlab.com/sevncodes/sevn.git}{https://gitlab.com/sevncodes/sevn.git}}. \cosmorate{} is publicly available on GitLab at \href{https://gitlab.com/Filippo.santoliquido/cosmo_rate_public}{this link}\footnote{\href{https://gitlab.com/Filippo.santoliquido/cosmo_rate_public}{https://gitlab.com/Filippo.santoliquido/cosmo\_rate\_public}}. Further data and codes will be shared on reasonable request to the corresponding authors.



\bibliographystyle{mnras}
\bibliography{santoliquido4} % if your bibtex file is called example.bib

%\begin{thebibliography}{99}
%\end{thebibliography}

%%%%%%%%%%%%%%%%%%%%%%%%%%%%%%%%%%%%%%%%%%%%%%%%%%

%%%%%%%%%%%%%%%%% APPENDICES %%%%%%%%%%%%%%%%%%%%%

\appendix

\section{Comparison sample of BBHs from Pop. I--II stars}
\label{sec:appendix}

In Figures~\ref{fig:mass0}--\ref{fig:mass1_dist}, we compare the masses of Pop.~III BBHs with those of Pop.~I-II BBHs. The latter are the fiducial model presented in \cite{iorio22}. Here, we briefly summarise their main features but we refer to \cite{iorio22} for more details. We simulate 5M binary star systems\footnote{The fiducial model by \cite{iorio22} only contains 1M binary star systems for each metallicity. Here, we  rerun the same model with a $5\times{}$ higher statistics, to filter out stochastic fluctuations.} for each of the following 15 metallicities:  $Z=10^{-4}$, $2\times10^{-4}$, $4\times10^{-4}$, $6\times10^{-4}$, $8\times10^{-4}$, $10^{-3}$, $2\times10^{-3}$, $4\times10^{-3}$, $6\times10^{-3}$, $8\times10^{-3}$, $10^{-2}$, $1.4\times10^{-2}$, $1.7\times10^{-2}$, $2\times10^{-2}$, $3\times10^{-2}$. The total number of simulated binary systems is thus 75M, ensuring that stochastic fluctuations are not important \citep{iorio22}.

The set-up of these simulations is the same as model KRO1, apart from the IMF mass range. In fact, we  randomly  draw the initial ZAMS mass of primary stars from a \citetalias{kroupa2001} IMF  with $M_\mathrm{ZAMS,1}\in \ [5,150] \ \mathrm{M}_\odot$ instead of $[5,\,{}550]$~M$_\odot$. We randomly select the masses of secondary stars 
assuming the distribution of mass ratios from \cite{sana2012} with a lower mass limit of $M_{\rm ZAMS,2}=2.2$~M$_\odot$. The initial orbital periods  and eccentricities  are also  generated according to the distributions  by \cite{sana2012}. The set-up of \sevn{} is the same as we describe in Section~\ref{sec:sevn}. 

From these simulations we extract our catalogues of Pop~I and II BBH mergers, which we use as input conditions for \cosmorate{}. We calculate the merger rate density of these metal-rich BBHs in the same way as described in Section~\ref{sec:cosmorate}, and in particular in Equation~\ref{eq:mrd}. To calculate $\mathcal{S}(z',Z)=\psi{}(z')\,{}p(z',Z)$, in this case we use 
\begin{equation}
\psi{}(z)= a\,{}\frac{(1+z)^{b}}{1+[(1+z)/c]^{d}}~~[\mathrm{M}_\odot~\mathrm{yr}^{-1}~\mathrm{Mpc}^{-3}],
\label{eq:madau2017}
\end{equation}
where $a=0.01$ M$_\odot$~Mpc$^{-3}$ yr$^{-1}$ (for a \citetalias{kroupa2001} IMF), $b = 2.6$, $c = 3.2$ and $d = 6.2$, 
from \cite{madau2017}.

We also assume an average metallicity evolution from \cite{madau2017}:
\begin{equation}
\label{eq:pdf2}
p(z', Z) = \frac{1}{\sqrt{2 \pi\,{}\sigma_{\rm Z}^2}}\,{} \exp\left\{{-\,{} \frac{\left[\log{(Z(z')/{\rm Z}_\odot)} - \langle{}\log{Z(z')/Z_\odot}\rangle{}\right]^2}{2\,{}\sigma_{\rm Z}^2}}\right\},
\end{equation}
where $\langle{}\log{Z(z')/Z_\odot}\rangle{}=\log{\langle{}Z(z')/Z_\odot\rangle{}}-\ln{(10)}\,{}\sigma_{\rm Z}^2/2$  and $\sigma_Z = 0.2$ \citep{bouffanais2021b}. Finally, in the calculation of the total initial stellar mass $\mathcal{M}_{\rm TOT}$, we introduce a term 
$\mathcal{M}_{\rm TOT}=M_{\rm sim}/f_{\rm IMF}$, where $M_{\rm sim}$ is the total initial simulated stellar mass and $f_{\rm IMF}=0.285$, to account for the fact that we simulate only stars with $M_\mathrm{ZAMS,1}>5$ M$_\odot$ and $M_\mathrm{ZAMS,2}>2.2$ M$_\odot$,  but we expect the \citetalias{kroupa2001} IMF to extend down to 0.1 M$_\odot$.

\section{Impact of mass accretion and common envelope efficiency}\label{app:mass_transfer}

%\textcolor{red}{\textbf{
The models shown in the main text assume mass-accretion efficiency $f_\mathrm{MT}=0.5$ for a non-degenerate accretor, and common-envelope efficiency $\alpha=1$. Here, we show the results of some additional models, in which we vary both $f_{\rm MT}$ and $\alpha$. In particular, we fix the initial binary orbital parameters to model LOG1 (Table~\ref{tab:IC}) and the star-formation history to \citetalias{hartwig2022}, and we explore the cases with $f_\mathrm{MT}=0.1,$ 0.5, 1.0 for a non-degenerate accretor, and $\alpha{}=0.5,$ 1, 3. The model with $\alpha=1$ and $f_\mathrm{MT}=1$ is the same as model LOG1 in Fig.~\ref{fig:mrd_asloth_uncrt}. %}}

\begin{figure}
    \centering
    \includegraphics[width = \columnwidth]{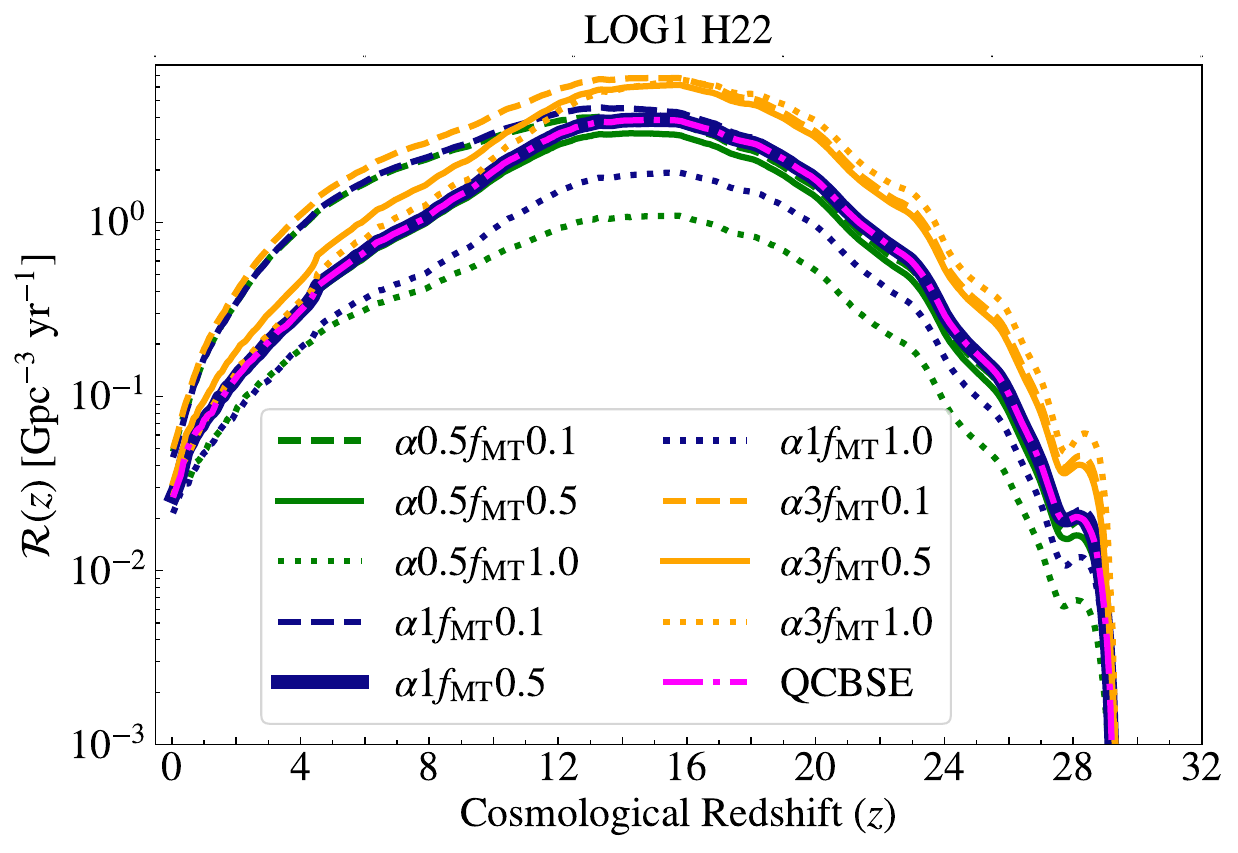}
    \caption{%\textcolor{red}{\bf 
    BBH Merger rate density evolution assuming the star formation rate history from \protect\citetalias{hartwig2022} and the initial binary orbital parameters as in model LOG1 (Table~\ref{tab:IC}). Dashed, solid and dotted lines refer to models with mass-accretion efficiency $f_{\rm MT}=0.1,$ 0.5, and 1, respectively. Green, blue and yellow lines refer to models with $\alpha=0.5,$ 1 and 3, respectively. The thick solid  blue line (with $f_{\rm MT}=0.5$ and $\alpha=1$) is the same as model LOG1 (solid blue line) in Fig.~\ref{fig:mrd_asloth_uncrt}. Finally, the dot-dashed magenta line (model QCBSE) is the same as the solid blue line but adopts the mass-transfer stability criteria by \protect\cite{hurley2002}. %}
    }
    \label{fig:rates_app}
\end{figure}

%\textcolor{red}{\textbf{ 
A low value of $f_{\rm MT}$ (highly non-conservative mass transfer) leads to higher merger rates at lower redshift, and is almost insensitive to the choice of $\alpha{}$ (Fig.~\ref{fig:rates_app}). This happens because the mass of the secondary star does not grow much during the first mass-transfer episode; when the primary star becomes a BH, the mass ratio between the companion star and the BH is never large enough to trigger a common envelope. Moreover, the binary system loses angular momentum because of mass loss and its semi-major axis shrinks: the progenitor binary star  steadily evolves via stable mass transfer and the resulting BBH merges with a long delay time. %}}

%when the primary star becomes a BH, the mass ratio between the companion star and the BH is never large enough to trigger a common envelope, and the system steadily evolves via stable mass transfer with long delay time.

%\textcolor{red}{{\bf 
In contrast, relatively large values of $f_{\rm MT}$ lead to an efficient mass growth of the secondary star, increasing the chances that the %BH-star 
system undergoes a common envelope episode (Fig.~\ref{fig:rates_app}). Larger values of $\alpha$ favour the ejection of the envelope and permit the survival of the binary system, whereas lower values of $\alpha$ trigger collisions between the star and the BH (or between the two progenitor stars), reducing the BBH merger rate density. %}}

\begin{figure*}
    \centering
    \includegraphics[width = \textwidth]{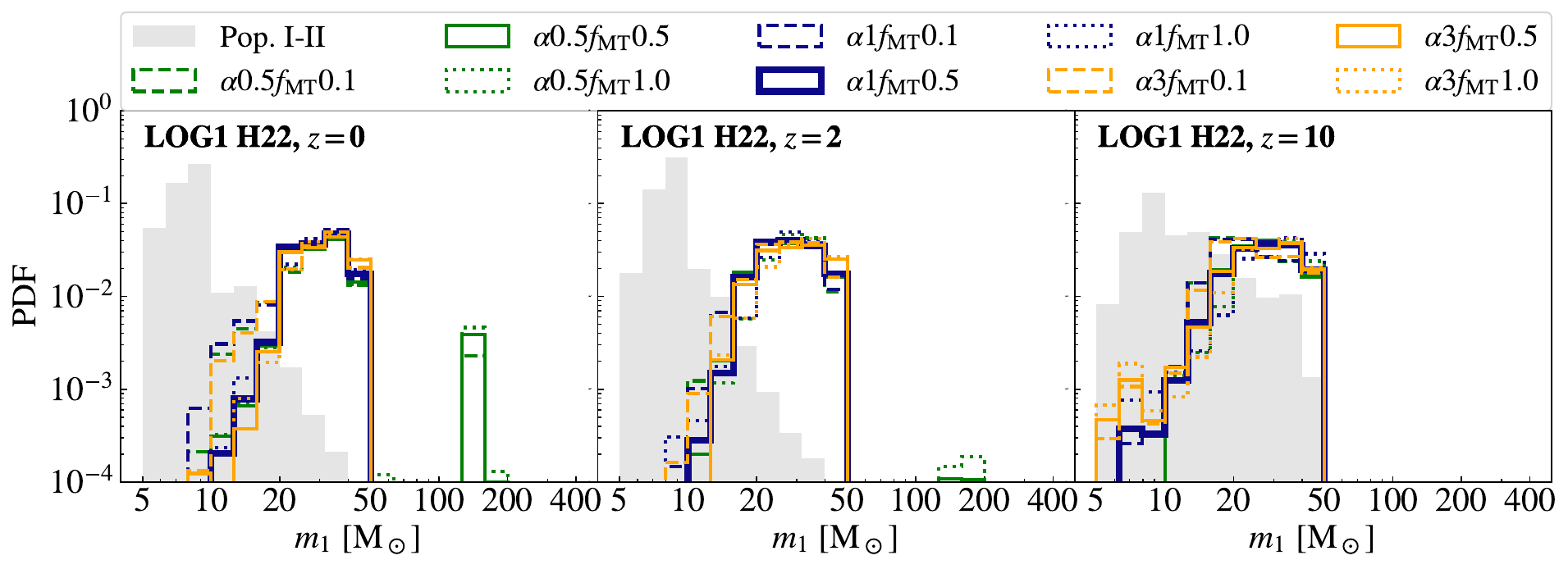}
    \caption{%\textcolor{red}{\bf 
    Primary BH mass distribution of Pop.~III BBHs merging at redshift $z=0,$ (left), 2 (middle), and 10 (right) assuming the star formation history from \protect\citetalias{hartwig2022} and the initial binary orbital parameters as in model LOG1 (Table~\ref{tab:IC}). Dashed, solid and dotted lines refer to models with mass-accretion efficiency $f_{\rm MT}=0.1,$ 0.5, and 1, respectively. Green, blue and yellow lines refer to models with $\alpha=0.5,$ 1 and 3, respectively. The grey shaded histogram shows the distribution of Pop.~I-II BHs for comparison. The model QCBSE (magenta line in Fig.~\ref{fig:rates_app}) is not shown here, because it perfectly overlaps with model LOG1 (thick solid blue line). %} % ADD CASE QCBSE?}
    }
    \label{fig:mass_app}
\end{figure*}

%\textcolor{red}{\textbf{
Such differences  have a mild impact on the distribution of the primary mass (Fig.~\ref{fig:mass_app}). BBHs with primary mass above the gap have a higher merger efficiency in the simulations with $\alpha=0.5$ at low redshift. These systems evolve via channel III: their orbital separation shrinks more efficiently with $\alpha=0.5$, allowing them to merge even if they have large initial orbital periods. %}}

%\textcolor{red}{\textbf{
Figure~\ref{fig:rates_app} also shows the comparison between our fiducial model (LOG1) and a simulation (QCBSE) in which we use the same mass-transfer stability criteria as \cite{hurley2002}. The main difference between LOG1 and QCBSE is that the former assumes that mass transfer is always stable for donor stars in the main sequence and Hertzsprung-gap phase, while the latter allows for mass transfer to become unstable in these early evolutionary phases. We find almost no difference between LOG1 and QCBSE,  because the progenitors of most of our successful BBH mergers undergo the first mass transfer episode when the primary star is  a post-Hertzsprung gap object. %}}

%Increasing the mass-accretion efficiency parameter from $f_{\rm MT}=0.1$ to  $f_{\rm MT}=1.0$ (conservative mass transfer) tends to lower the merger rate density and to shift its peak to a higher redshift. A conservative mass transfer increases the mass of the secondary star more efficiently during the first mass-transfer episode. This leads to a larger donor-to-accretor mass ratio during  later mass-transfer episodes, when the secondary star becomes the donor, thus increasing the chances that the star--BH system undergoes common envelope.
%and thus stabilises the system against common envelope, decreasing the overall merger rate density.}}

%\textcolor{red}{{Increasing the $\alpha$ parameter of common-envelope evolution from $\alpha=0.5$ to 3 produces an increase of the merger rate density, especially for high values of $f_\mathrm{MT}$}}

%%%%%%%%%%%%%FIGURE%%%%%%%%%%%%%%%%%
%%%%%%%%%%%%%%%%%%%%%%%

%%%%%%%%%%%%%FIGURE%%%%%%%%%%%%%%%%%
%%%%%%%%%%%%%%%%%%%%%%%

%%%%%%%%%%%%%%%%%%%%%%%%%%%%%%%%%%%%%%%%%%%%%%%%%%

% Don't change these lines
\bsp	% typesetting comment
\label{lastpage}
\end{document}